\documentclass[reprint,aps,prl,twocolumn,superscriptaddress,nobibnotes]{revtex4-2}
\usepackage[english]{babel}
\usepackage[utf8]{inputenc}

\usepackage{graphicx,epsfig,xcolor}

\usepackage{natbib}
\usepackage{amsfonts}
\usepackage{amsmath}
\usepackage{amssymb}
\usepackage{latexsym}
\usepackage{cancel}
\usepackage{hyperref}
\usepackage{placeins}
\onecolumngrid
\usepackage{polski}
\usepackage{braket}
\usepackage{soul}

\newcommand*{\Comb}[2]{{}^{#1}C_{#2}}
\newcommand{\Tr}{\mathrm{Tr}}
\def\ket|#1>{| #1 \rangle}
\def\bra<#1|{\langle #1 |}
\def\<{\langle}
\def\>{\rangle}
\def\{{\lbrace}
\def\}{\rbrace}
\def\({\left(}
\def\){\right)}
\def\beq{\begin{equation}}
\def\eeq{\end{equation}}

\newcommand{\rmvd}[1]{}

\begin{document}

\title{Unveiling Eigenstate Thermalization for Non-Hermitian systems}

\author{Sudipto Singha Roy}
\thanks{These two authors contributed equally}
\affiliation{Pitaevskii BEC Center, CNR-INO and Dipartimento di Fisica, Universit\`a di Trento, Via Sommarive 14, Trento, I-38123, Italy}
\affiliation{INFN-TIFPA, Trento Institute for Fundamental Physics and Applications, Trento, Italy}
\affiliation{Department of Physics, Indian Institute of Technology (ISM) Dhanbad, IN-826004, Dhanbad, India}
\author{Soumik Bandyopadhyay}
\thanks{These two authors contributed equally}
\affiliation{Pitaevskii BEC Center, CNR-INO and Dipartimento di Fisica, Universit\`a di Trento, Via Sommarive 14, Trento, I-38123, Italy}
\affiliation{INFN-TIFPA, Trento Institute for Fundamental Physics and Applications, Trento, Italy}
\author{Ricardo Costa de Almeida}
\affiliation{Pitaevskii BEC Center, CNR-INO and Dipartimento di Fisica, Universit\`a di Trento, Via Sommarive 14, Trento, I-38123, Italy}
\affiliation{INFN-TIFPA, Trento Institute for Fundamental Physics and Applications, Trento, Italy}
\author{Philipp Hauke}
\email{philipp.hauke@unitn.it }
\affiliation{Pitaevskii BEC Center, CNR-INO and Dipartimento di Fisica, Universit\`a di Trento, Via Sommarive 14, Trento, I-38123, Italy}
\affiliation{INFN-TIFPA, Trento Institute for Fundamental Physics and Applications, Trento, Italy}

\begin{abstract}
The eigenstate thermalization hypothesis (ETH) has been highly influential in explaining thermodynamic behavior of closed quantum systems. 
As of yet, it is unclear whether and how the ETH applies to non-Hermitian systems. 
Here, we introduce a framework that extends the ETH to non-Hermitian systems, within which expectation values of local operators reproduce statistical and scaling predictions known from Hermitian ETH.  
We illustrate the validity of the framework on non-Hermitian random-matrix and 
Sachdev-Ye-Kitaev models. Further, we show numerically how the static ETH predictions 
become imprinted onto the dynamics of local observables. Finally, we present a 
prescription for observing both ETH-obeying and ETH-violating regimes in an 
optical-lattice experiment that implements a disordered interacting Hatano-Nelson 
model.
Our results generalize the celebrated ETH to the non-Hermitian setting, and they show 
how it affects the system dynamics, and how the salient signatures can be observed in 
present-day cold-atom experiments.
\end{abstract}

\date{\today}

\maketitle

{\it Introduction---}
The eigenstate thermalization hypothesis (ETH)~\cite{ETH_ansatz1, ETH_ansatz2, ETH_ansatz3, ETH_ansatz4, ETH_ansatz5, ETH_ansatz7,ETH_ansatz8} plays an instrumental role in understanding the thermalization of closed quantum systems.  
Its central claim is that  expectation  values of generic local observables of nonintegrable models follow a simple form that depends only on smooth functions of energy and density of states. 
Over the past years, tremendous efforts have been put into understanding the applicability of the ETH, for many models
~\cite{ETH_local1,ETH_local2,ETH_local3,ETH_local4,ETH_local5,ETH_local6,ETH_SYK1,ETH_SYK2,ETH_CFT1, ETH_CFT2,ETH_longrange1,ETH_longrange2,ETH_longrange3,ETH_ansatz9,ETH_generalization1,ETH_generalization2,
ETH_generalization3,ETH_generalization4, ETH_generalization5,ETH_generalization6,ETH_generalization7,ETH_generalization8,pretherm1,pretherm2,Zhou,scars,SYK5}. 
However, with the exception of Ref.~\cite{ETH_nh}, none of these cover models described by non-Hermitian Hamiltonians. 
Non-Hermitian systems show intriguing phenomena not present in their Hermitian counterparts, including complex eigenspectra~\cite{Blender,Christodoulides, Ueda,Rotter1, Jin, Biella,Jo},  exceptional points~\cite{Ueda, Nori,Zhen,Miri,Franchi}, coalescence  of eigenstates~\cite{Graefe,Chan}, the non-Hermitian skin effect~\cite{skin_effects1,skin_effects2,skin_effects3,skin_effects4,skin_effects5}, and non-Hermitian linear response~\cite{Zhai,Kevin}, and they have been proposed as platforms for quantum-enhanced sensing~\cite{nh_sensing1,nh_sensing2,nh_sensing3,nh_sensing4,nh_sensing5,nh_sensing6}.
The widespread theoretical effort for achieving a deeper understanding is going hand in hand with a strong advance in the experimental control over non-Hermitian systems~\cite{Luo,Choi, Chan,Franchi, Weith,Shen, DeCarlo}.  
Despite their rising importance, it remains an outstanding challenge to understand the emergence of ETH in the context of non-Hermitian systems.  

\begin{figure}
\includegraphics[width=\linewidth]{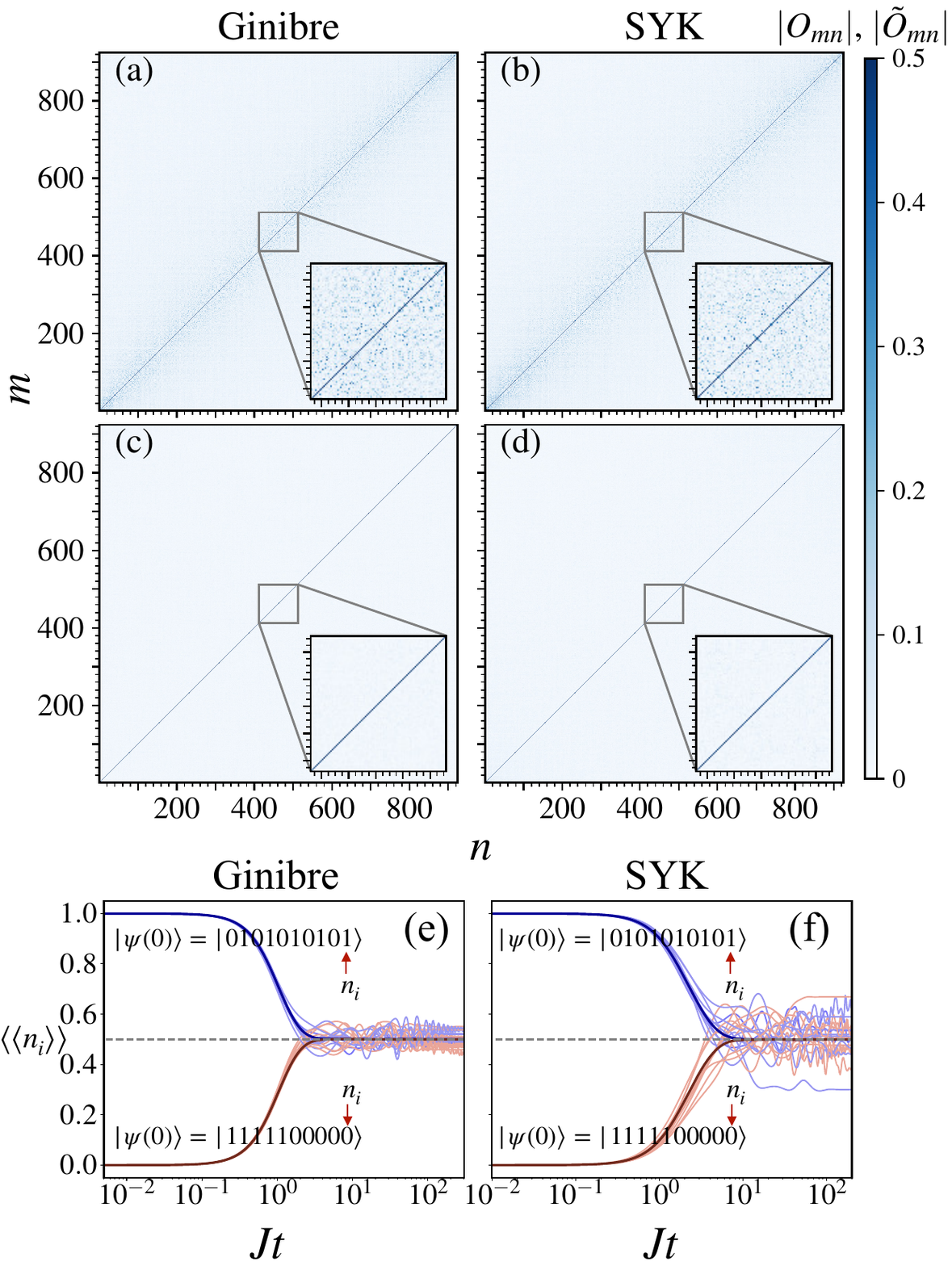}
\vspace*{-0.75cm}
 \caption{
The evolution under a generic non-Hermitian Hamiltonian $H_{\mathrm{nh}}$ ($H_{\mathrm{nh}}^\dagger$) naturally leads to measuring the expectation value of an operator $\hat{O}$ in right (left) eigenvectors.
(a,b) $|\langle R_m|\hat{O}|R_n\rangle|$ for a non-Hermitian random matrix (a) and SYK model (b), for a single realization and system size $N=12$. Unlike the behavior obtained for the biorthogonal basis~\cite{sup_mat}, diagonal elements ($m=n$) show the signatures of ETH. However, the bare off-diagonal elements ($m\neq n$) exhibit significant nonzero contributions 
   (zooms). 
   (c,d) Once we correct for the off-diagonal contribution that comes solely from the nonorthogonality of eigenvectors (see Eq.~(\ref{eq:ETH_modified_substraction})), the  behavior is similar to the Hermitian counterpart and in agreement with ETH. (e,f) Thermalization dynamics of local number operator under non-Hermitian evolution, for (e) the random matrix $H_{\mathrm{nh}}$ obtained from the Ginibre ensemble and (f) the SYK model. Already after $Jt\sim 5$, the local number operator fluctuates around its mean value $\overline{n}_{i}=0.5$ (dashed grey line),  yielding a long-time average that closely aligns with the non-Hermitian ETH predictions derived in (c,d). For details, see 
  \cite{sup_mat}.
 }
  \label{fig:ROR_Ginibre}
\end{figure}

In this Letter, we establish a framework for the ETH in non-Hermitian systems.
Specifically, we construct a modified ansatz that enables us to generalize the celebrated Hermitian ETH ansatz.
It accounts for the non-orthogonality of eigenstates of non-Hermitian Hamiltonians, and enables us to correctly describe both diagonal and off-diagonal matrix elements of local expectation values (see Fig.~\ref{fig:ROR_Ginibre}).
As a main application of ETH is to predict quantum out-of-equilibrium dynamics, we argue that the basis of right eigenvector has to be used, in contrast to the biorthogonal basis, which leads to expectation values incompatible with ETH~\cite{ETH_nh}.
We confirm our framework through detailed numerics on non-Hermitian random matrix ensembles and non-Hermitian versions of the Sachdev-Ye-Kitaev (SYK) model. For both models, we explicitly demonstrate that the long-time dynamics of local observables converge to their mean values, providing strong support for the non-Hermitian ETH framework. Additionally, we consider the interacting Hatano-Nelson model with nonreciprocal hopping and disorder~\cite{Hatano1, Hatano2, Hatano3}.  
We demonstrate the applicability (breakdown) of non-Hermitian ETH in the weak (strong) disorder limit, where the system  thermalizes (localizes), and provide a feasible prescription to observe this physics in present-day cold-atom experiments 
\cite{Schreiber_2015, choi_2016, Bordia_2017, Luschen_2017, Rispoli_2019}.Our work thus provides a complete recipe for extending ETH into the non-Hermitian paradigm and certifies its validity  through the long-time behavior of local observables.


{\it Background---} The ETH prediction for Hermitian random matrix theory (RMT) is~\cite{ETH_ansatz7} 
\begin{equation}
O_{mn}=\langle m|\hat{O}|n \rangle=\overline{O} \delta_{mn}+
\tilde{d}_{mn} \sqrt{\frac{\overline{O^2}}{\mathcal{D}}},
\label{eq:ETH_RMT}
\end{equation}
where $\overline{O}$ is the mean of the observable, $\mathcal{D}$ is the Hilbert space dimension, and $d_{mn}$ and  $\tilde{d}_{mn}$ are random variables with unit variance and vanishing mean. 

It is \emph{a priori} not obvious how to apply these relations to non-Hermitian systems. In particular, one unusual property of non-Hermitian Hamiltonians is that their left and right eigenvectors are orthogonal but not necessarily the Hermitian conjugate of each other. 
In formulas, a non-Hermitian Hamiltonian can be diagonalized as 
\begin{equation}
H_{\mathrm{nh}}=\sum_m \varepsilon_m |{R_m}\rangle\langle{L_m}|\,,
\end{equation}
with $\langle L_m| R_n \rangle =\delta_{mn}$, but $\langle L_m|\neq (|R_m\rangle)^\dagger$ is possible. 
This behavior of $H_{\mathrm{nh}}$ has led to diverging attempts when trying to analyze its properties: it is currently unclear whether one should choose the state of interest as $\rho^{RR}_m=|R_m\rangle \langle R_m|$, as $\rho_m^{LL}=|L_m\rangle \langle L_m|$,  or, using the biorthogonal basis states, as $\rho_m^{RL}=|R_m\rangle \langle L_m|$~\cite{Bardarson, Ryu,Guo_2021, Takato_2020,Zhenming_2024}. 
In particular, when defining the ``ground state" of $H_{\mathrm{nh}}$ as $\rho^{RL}_{0}=|R_0\rangle \langle L_0|$, results such as complex entanglement entropy appear~\cite{Ryu, Modak}, which lack clear physical interpretation.

To a considerable part, the significance of the ETH stems from its relevance to time evolution of quantum systems. As we show in the Supplemental Material (SM)~\cite{sup_mat}, expectation values of observables, $\Tr(\rho(t)\hat{O})$, will be determined by the matrix overlap elements $\langle R_n|\hat{O}|R_m\rangle $~\cite{time_evol_righteigen}.
The left eigenvectors enter only through the overlap with the initial state. Mixed matrix elements of the form $\langle R_n|\hat{O}|L_m\rangle $ do not appear. 
Motivated by this physical interpretation, we will use the basis given by the right eigenvectors.

{\it Formalism and non-Hermitian RMT analysis---}The formalism we introduce here is best illustrated through a non-Hermitian RMT analysis.  Statistical properties of typical chaotic non-Hermitian systems can be modeled by drawing the elements of $H_{\mathrm{nh}}$ from the complex Ginibre ensemble~\cite{Ginibre_non_Herm1, Ginibre_non_Herm2, Ginibre_non_Herm3, Ginibre_non_Herm4} (see SM~\cite{sup_mat} for details). 
As a local observable of interest, we choose one that is similar to what we will use to probe the non-Hermitian  SYK model below. Specifically, we choose the single-mode number operator $\hat{O}=\hat{n}_i=c^{\dagger}_i c_i$, where $c^\dagger_i$ ($c_i$) is the creation  (annihilation) operator for the $i$'th fermionic mode.
Throughout, we represent Hamiltonians, operators, and states of interest using the standard fermionic field operators.  
As a result, even though the right eigenvectors are not mutually orthogonal, when computing physical observables the fermionic statistics is respected. Additionally, as we will restrict our studies of the SYK model to the half-filling sector, we choose  $H_{\mathrm{nh}}$ as a  $\mathcal{D} \times \mathcal{D}$ random matrix with $\mathcal{D}=\Comb{N}{N/2}$.  Moreover,  throughout this work, we consider the right eigenvectors to be properly normalized, i.e.,  $\langle R_m|R_m\rangle=1$.  We now provide a systematic analysis of both diagonal and off-diagonal terms of $\langle R_m|\hat{O}|R_n\rangle$. 

\begin{figure}
 \includegraphics[width=0.98\linewidth]{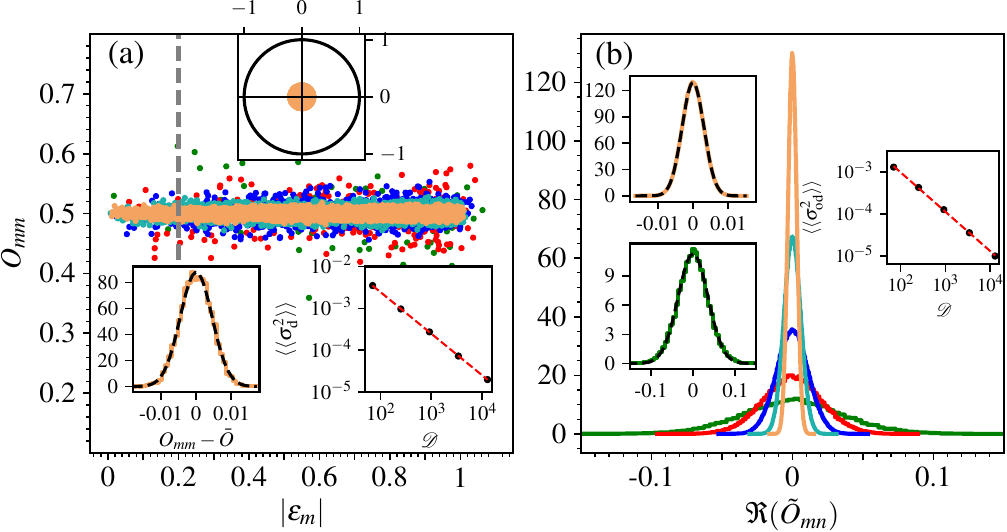}
  \caption{Signature of ETH in a random non-Hermitian Hamiltonian from the Ginibre ensemble. 
  (a) Diagonal elements of the local observable $\hat{O}$ computed in the right eigenvectors, $|O_{mm}|= |\langle R_m|\hat{O}|R_m\rangle|$, against $|\epsilon_m|$, for system sizes $N=8,10,12,14, 16$ (green, red, blue, sea green, orange)  and a single random realization. 
  With increasing size, fluctuations are suppressed and $O_{mm}$ tend to converge to $\overline{O}=0.5$. 
 Bottom left inset: histogram of  $[\overline{O}-O_{mm}]$ for eigenstates within $|\epsilon_m|\leq 0.2$ (top inset), 
  for $N=16$ and for $20$ random realizations. The profile is fitted well with a Gaussian distribution (black dashed). 
  Bottom right inset: fluctuations, as measured by the variance $\langle \langle \sigma_{\mathrm{d}}^2 \rangle \rangle$ of $[\overline{O}-O_{mm}]$, shows a perfect linear decay (having slope $\approx 0.99$) with the Hilbert space dimension $\mathcal{D}$ in logarithmic scale, thus confirming an exponential decrease with system size (data for $5000$, $1000$, $500$, $200$, and $20$ realizations from $N=8$ to $N=16$). 
(b) Histogram of the real part of the off-diagonal terms $\widetilde{O}_{mn}=O_{mn}-\langle R_m|R_n\rangle \overline{O}$, $m \neq n$, again obtained for energy states with $|\epsilon_m|\leq 0.2$ but for different system sizes, $N=8,10,12,14, 16$ (green, red, blue, sea green, orange), with $5000$, $1000$, $500$, $200$, $20$ random realizations, respectively. The profiles can be accurately fitted by a Gaussian distribution, as exemplified for $N=8$ (left bottom inset) and $N=16$ (left top inset). As for the diagonal case, the variance of the off-diagonal terms $\widetilde{O}_{mn}$,   $\langle \langle \sigma_{\mathrm{od}}^2 \rangle \rangle$, also exhibits exponential decay with system size (right inset). In this case the fitted slope $\approx 0.94$.}
  \label{fig:ROR_corrected_Ginibre}
\end{figure}

{\it Diagonal terms---} 
Figure~\ref{fig:ROR_Ginibre}(b) illustrates the overall behavior of $O_{mn}=|\langle R_m|\hat{O}|R_n\rangle|$  for $N=12$ and one single instance of random realization. 
For a more quantitative analysis, Fig.~\ref{fig:ROR_corrected_Ginibre}(a) reports the diagonal terms ($m=n$) for different system sizes $N=8, 10, 12, 14, 16$ against 
$|\epsilon_m|$. 
As this data shows, fluctuations around the mean $\overline{O}=\frac{1}{\mathcal{D}}\sum_p \langle p|\hat{O}|p\rangle$ are  
strongly suppressed with system size. Here, $|p\rangle=|p_1,\dots,p_N\rangle$ are the basis states of the fermionic  Fock space. Two insets provide further quantitative support:  
in the bottom left inset of Fig.~\ref{fig:ROR_corrected_Ginibre}(a), we plot the histogram of $[O_{mm}-\overline{O}]$ obtained using all the eigenstates lying within the energy disk of radius $|\epsilon_m|=0.2$ (indicated in the top inset), for $20$ random realizations of the model and $N=16$ (see also SM~\cite{sup_mat}). The data is well fitted by a Gaussian profile (black dashed line).  Moreover, in the right inset  we display the variance of $[O_{mm}-\overline{O}]$, $\langle \langle \sigma_{\mathrm{d}}^2 \rangle \rangle$, against Hilbert space dimension $\mathcal{D}$, exhibiting a clear exponential decay with system size, a property also found in nonintegrable Hermitian systems~\cite{ETH_local3, ETH_off_diagonal1,ETH_off_diagonal3}. 

This behavior remains thus close  to the  RMT prediction for Hermitian systems, i.e., for a single instance of random realization we find 
\begin{eqnarray}
O_{mm}=\langle R_m|\hat{O}|R_m\rangle=\overline{O}+ Q_{mm} \sqrt{\frac{\overline{O^2}}{\mathcal{D}}}, 
\label{eq:ETH_Ginibre_diag}
\end{eqnarray}
where the second term describes the exponentially decreasing fluctuations of $\langle R_m|\hat{O}|R_m\rangle$ governed by independent random numbers $Q_{mm}$.

{\it Off-diagonal terms---} 
As Fig.~\ref{fig:ROR_Ginibre}(b)  illustrates, the off-diagonal elements $|\langle R_m|\hat{O}|R_n\rangle|_{m\neq n}$ have significantly less structure than the elements evaluated in the biorthogonal basis (see~\cite{sup_mat}). At first sight, this result is nevertheless still not compatible with ETH, as the inset of Fig.~\ref{fig:ROR_Ginibre}(b) shows, since (i) there is some remaining structure and (ii) the off-diagonal terms do not get suppressed with Hilbert space dimension even at moderately large system size. The reason lies in the unusual orthogonality relations of non-Hermitian eigenstates. In what follows, we first detail the physical reasoning, based on which we then present the necessary modification of Eq.~(\ref{eq:ETH_RMT}) that restores the notion of ETH for the off-diagonal terms.

As mentioned before, the right eigenvectors of a non-Hermitian system can be nonorthogonal, $\langle R_m|R_n\rangle \neq 0$ for $m\neq n$. However, one can always isolate the (normalized) component of $|R_{n}\rangle$ orthogonal to  $|R_m\rangle$, $|R_{n \perp m}\rangle$, through the decomposition  
\begin{eqnarray}
|R_n\rangle=\alpha_{nm} |R_m\rangle+\beta_{nm} |R_{n\perp m}\rangle, 
\label{eqn:decomposition}
\end{eqnarray}
where $\alpha_{nm}=\langle R_m|R_n\rangle$ and $\beta_{nm}=\langle R_{n \perp m}|R_n\rangle$. 
Using this relation, we get 
\begin{align}
\label{eq:oddecomposed}
\langle R_m|\hat{O}|R_n\rangle_{m\neq n}=\alpha_{nm}{O}_{mm}  
+ \beta_{nm} \langle R_m|\hat{O}|R_{n \perp m}\rangle.
\end{align}
Since $|R_{m}\rangle$ and $|R_{n \perp m}\rangle$ are orthonormal random vectors, we can expect the second term to behave as per the prediction from the Hermitian ETH, Eq.~\eqref{eq:ETH_RMT}, i.e., to be of the form $\beta_{nm}\tilde{Q}_{mn} \sqrt{\frac{\overline{O^2}}{\mathcal{D}}}$, with $\tilde{Q}_{mn}$ a random number. 
Using the ETH for the diagonal elements as established above, the first term will behave as 
$ \alpha_{nm} \overline{O}+\alpha_{nm} Q_{mm}\sqrt{\frac{\overline{O^2}}{\mathcal{D}}}$. 
The randomly fluctuating parts can be collected into a term $S_{mn} \sqrt{\frac{\overline{O^2}}{\mathcal{D}}}$, where $S_{mn}=\alpha_{nm}Q_{mm}+\beta_{nm} \tilde{Q}_{mn}$ is again a random number. This term would obey the ETH for off-diagonal elements, as prescribed for the Hermitian ETH, Eq.~\eqref{eq:ETH_RMT}. 
There remains, however, a further contribution to the off-diagonal elements that derives from the non-orthogonality of eigenvectors, $\alpha_{nm} \overline{O}$. 
As generically in non-Hermitian systems $\alpha_{nm}=\langle R_m|R_n\rangle \neq 0$, even at moderately large system size $\langle R_m|\hat{O}|R_n\rangle$ can thus still have a significant nonzero value. We hence need to suitably modify the ETH relation. 

Summarizing the above considerations, we can thus conjecture the generalized version of the ETH ansatz for a non-Hermitian random matrix: 
\noindent{\it The matrix elements of any local observable, $\hat{O}$, obeying the ETH ansatz for  any generic normalized basis $\{|R_m\rangle\}$ of a non-Hermitian random matrix Hamiltonian should behave as }
\begin{align}
O_{mm}&=\langle R_m|\hat{O}|R_m\rangle=\overline{O}+ Q_{mm} \sqrt{\frac{\overline{O^2}}{\mathcal{D}}}, \label{eq:ETH_modified_Omm_RMT}\\
O_{mn}&=\langle R_m|\hat{O}|R_n\rangle_{m \neq n}=\langle R_m|R_n\rangle \overline{O}+S_{mn} \sqrt{\frac{\overline{O^2}}{\mathcal{D}}}.
\label{eq:ETH_modified_Omn_RMT}
\end{align}
In case of Hermitian systems,  $\langle R_m|R_n\rangle=\delta_{mn}$ and we recover the familiar  prediction of Eq.~\eqref{eq:ETH_RMT} as a special case. 
We can also conjecture that for a generic non-Hermitian system, $\overline{O}$ and $\sqrt{\frac{\overline{O^2}}{\mathcal{D}}}$ should be replaced by smooth functions of the (complex) average energy $\overline{E}$ and energy difference $\omega$.

We present the numerical results supporting Eqs.~(\ref{eq:ETH_modified_Omm_RMT}) and (\ref{eq:ETH_modified_Omn_RMT}) in Fig.~\ref{fig:ROR_Ginibre}(d), where we plot the absolute value of 
\begin{align}
\widetilde{O}_{mn}= 
\begin{cases}
    O_{mn}& \text{for}~m=n,\\
    O_{mn}-\langle R_m|R_n\rangle\overline{O}             & \text{for}~ m\neq n\,.
    \label{eq:ETH_modified_substraction}
\end{cases}
\end{align}
As the figure illustrates, the modification of the off-diagonal terms unveils a behavior of the local observable exactly analogous to the prediction for the  Hermitian systems.  Moreover, as the histograms of $\mathfrak{R}[\widetilde{O}_{mn}]_{n \neq m}$, plotted in Fig.~\ref{fig:ROR_corrected_Ginibre}(b), show, $\widetilde{O}_{mn}$ gets suppressed with increasing system size and approximates a Gaussian profile.  
Additionally, analogously to the Hermitian case~\cite{ETH_ansatz7,ratio_diag}, the ratio of the variance of diagonal ($O_{mm}$) and modified off-diagonal ($\widetilde{O}_{mn}$)  expectation values becomes $\frac{\langle \langle \sigma_{\mathrm{d}}^2\rangle \rangle}{\langle \langle \sigma_{\mathrm{od}}^2\rangle \rangle} \approx 2$.  
All these observations strongly corroborate the signature of ETH in the non-Hermitian random Hamiltonian $H_{\mathrm{nh}}$. 

To further substantiate this claim, we explicitly compute the thermalization dynamics of local observables for an initial state evolved under the non-Hermitian Ginibre Hamiltonian, Fig.~ \ref{fig:ROR_Ginibre}(e). Their behavior clearly and rapidly converges to the average value predicted by the proposed non-Hermitian ETH.

\begin{figure}
 \includegraphics[width=0.98\linewidth]{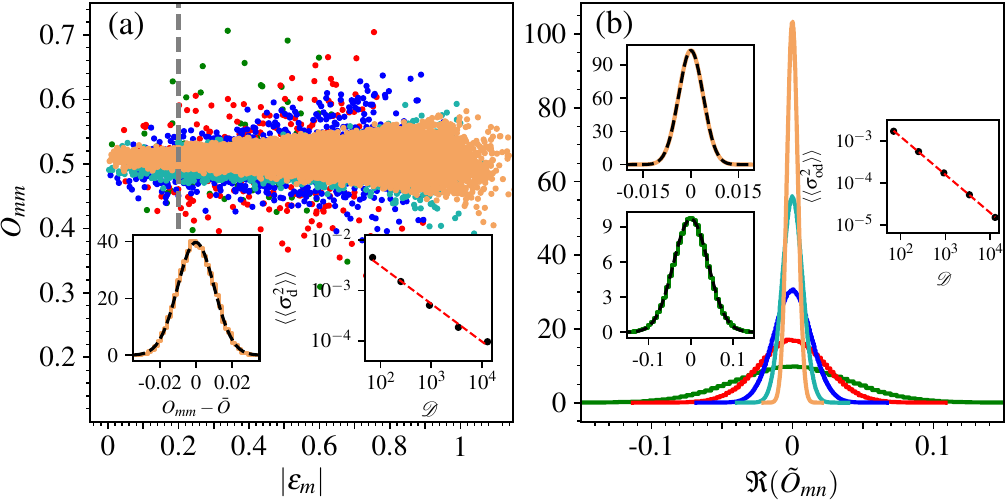}
  \caption{Signature of ETH in the non-Hermitian SYK model. (a) Diagonal elements $|\langle R_m|\hat{O}|R_m\rangle|$ of $\hat{O}=\hat{n}_i$, against $|\epsilon_m|$ for system sizes $N = 8, 10, 12, 14, 16$ (green, red, blue, sea green, orange) and for a single random realization. Similar to Fig.~\ref{fig:ROR_corrected_Ginibre}, fluctuations here are also suppressed as system size increases. However, unlike the random ensemble case, the spread around the mean is not uniform across the energy range. High-energy states show more deviation than the low-energy ones, which is a characteristic feature of the Hermitian SYK Hamiltonian. 
  Bottom left inset: The histogram of $[O_{mm}-\overline{O}]$ for eigenstates within the range $|\epsilon_m|\leq 0.2$ (for $N=16$ and $20$ random realizations of the Hamiltonian) is fitted well by a Gaussian curve (black dashed). The bottom right inset depicts linear decay (plotted in logarithmic scale) of  variance of diagonal term $\langle \langle \sigma_{\mathrm{d}}^2 \rangle \rangle$ with Hilbert space dimension $\mathcal{D}$ for different system sizes and random realizations of the Hamiltonian that are considered exactly the same as  in Fig.~\ref{fig:ROR_corrected_Ginibre}(a). 
  (b) Distribution of the $\mathfrak{R}[\widetilde{O}]_{m\neq n}$, 
  obtained for eigenenergies as in (a) (for system sizes $N=8, 10, 12, 14, 16$, and   marked by green, red, blue, sea green, and orange color, respectively). As the Gaussian fits shown for $N=8$ (bottom left inset) and $16$ (top left inset), and the exponential decay of the off-diagonal variance $\langle \langle \sigma_{\mathrm{od}}^2 \rangle \rangle$ (all the parameters considered for this are same as in Fig.~\ref{fig:ROR_corrected_Ginibre}(b)) with the Hilbert space dimension $\mathcal{D}$ highlight, by correctly modifying the off-diagonal description the non-Hermitian SYK model displays the signatures of ETH.}
  \label{fig:ROR_corrected_SYK}
\end{figure}

{\it Sachdev-Ye-Kitaev model---}In the final part of our Letter, we confirm the prediction given in Eqs.~(\ref{eq:ETH_modified_Omm_RMT}, \ref{eq:ETH_modified_Omn_RMT}) for the non-Hermitian counterpart of the SYK model~\cite{SYK1, SYK2, SYK3, SYK4, SYK5}. The SYK model consists of $N$ fermionic modes with  disordered all-to-all interactions, described by  (see also~\cite{SYK_nonh_other1, SYK_nonh_other2,SYK_nonh_other3})
\begin{eqnarray}
H^{\text{SYK}_4}_{\mathrm{nh}}=\frac{1}{(2N)^{\frac{3}{2} }}\sum_{i_1 i_2 j_1 j_2} J_{i_1i_2:j_1j_2} \hat{c}^\dagger_{i_1} \hat{c}^\dagger_{i_2} \hat{c}_{j_1} \hat{c}_{j_2}\,.
\label{eqn:SYK}
\end{eqnarray}
Here, we consider the variant with spinless complex fermions, with $\hat{c}_i$ as defined before. The real and imaginary parts of $J_{i_1 i_2:j_1 j_2}$ are independent and normally distributed. The non-Hermitian  version is obtained by choosing complex $J_{i_1i_2:j_1j_2}$ for the diagonal elements and imposing $J_{i_1i_2:j_1j_2}^* \neq J_{j_1j_2:i_1i_2}$ for the off-diagonals. The following behavior is generic also for other choices of complex interaction amplitudes, see~\cite{sup_mat} for details.

As for the random ensemble, we consider expectation values of the local observable $\hat{O}=\hat{n}_i$, see Fig.~\ref{fig:ROR_Ginibre}(c). 
Again, the diagonal elements for the non-Hermitian SYK model remain in accordance with its Hermitian counterpart and show the characteristics of the ETH. 
In the bare off-diagonal terms, strong fluctuations are prominent that can be seen in the zoomed inset. 
Subtracting the contribution coming solely from the nonorthogonality relation, i.e., employing Eq.~\eqref{eq:ETH_modified_substraction}, we again obtain clearly the signatures of ETH as well as the off-diagonal terms, see Fig.~\ref{fig:ROR_Ginibre}(e). 

For completeness, in Fig.~\ref{fig:ROR_corrected_SYK} we reproduce the analysis analogous to Fig.~\ref{fig:ROR_corrected_Ginibre}. 
As before, the diagonal elements $|\langle R_m|\hat{O}|R_m\rangle|$  converge to the mean value [Fig.~\ref{fig:ROR_corrected_SYK}(a)]; however, unlike the random ensemble case, the behavior is not uniform in $|\epsilon_m|$, with low-energy states (small $|\epsilon_m|$) converging faster. This behavior is in accordance with the one for the Hermitian case~\cite{ETH_SYK1}. 
To strengthen our numerical claim, in the inset (bottom left) we present the histogram of $[\langle R_m|\hat{O}|R_m \rangle-O(\overline{E})]$ for eigenstates within $|\epsilon_m|\leq 0.2$, where $\overline{E}$ is the average energy within the disk, using $N=16$ and 20 random realizations. The profile accurately coincides with a Gaussian distribution (black dashed).  A similar analysis  for the off-diagonal terms is shown in Fig.~\ref{fig:ROR_corrected_SYK}(b), where we plot the scaling of the real part of $\widetilde{O}_{mn}=O_{mn}-\langle R_m|R_n\rangle \overline{O}$, ${m \neq n}$  for system sizes $N=8, 10, 12, 14, 16$. Also, for this data, we find excellent agreement with a Gaussian  
and exponential suppression of the variance $\langle \langle \sigma_{\mathrm{od}}^2 \rangle \rangle$ with system size. 

As for the Ginibre ensemble, the quench dynamics of local observables under the non-Hermitian Hamiltonian $H^{\text{SYK}_4}_{\mathrm{nh}}$ quickly convergences toward the mean value, see Fig.~\ref{fig:ROR_Ginibre}(f). 
The stronger fluctuations derive from the non-uniform behavior of $O_{mm}$ with $|\epsilon_m|$ as shown in   Fig.~\ref{fig:ROR_corrected_SYK}(a).

{\it Hatano--Nelson model and experimental observability---}In SM~\cite{sup_mat}, we analyze the static and dynamic ETH of an experimentally feasible interacting non-Hermitian Hamiltonian, the disordered Hatano--Nelson model with non-reciprocal hopping. 
As our detailed analysis illustrates,
the long-time dynamics of local observables evidences both regimes where non-Hermitian ETH predictions hold (at weak disorder) and regimes (at strong disorder) where they are violated. 

Importantly, this model can be implemented in present-day cold-atom experiments. 
An optical lattice superimposed with a running wave can realize non-reciprocal hopping~\cite{Gong_2018, Tomita_2019}, to which interactions and disorder can be added via the use of dipolar atoms~\cite{Baier_2016, su_2023} and random AC-Stark shifts~\cite{Billy_2008, Roati_2008, Schreiber_2015, choi_2016, Bordia_2017, Luschen_2017, Rispoli_2019, Sauerwein_2023}, respectively. Key observables such as local occupation or imbalance are accessible via quantum-gas microscopy~\cite{Gross_2021,Schreiber_2015, choi_2016, Bordia_2017, Luschen_2017, Rispoli_2019}, and the salient features are visible already on timescales of $Jt\sim10$, thus providing a clear scenario for experimentally probing non-Hermitian ETH.


{\it Discussion.---}
In this Letter, we have generalized the notion of ETH to non-Hermitian systems and have confirmed the framework for the random matrix ensemble and a SYK Hamiltonian.
The framework has strong physical consequences: when non-Hermitian ETH holds, the dynamics of local observables converges to their mean values. 
The ability to induce controlled quench dynamics in quantum simulators based on cold atoms and other platforms thus opens the way for direct laboratory tests of the non-Hermitian ETH~\cite{nh_our2}.

In the future, it will be interesting to apply our formalism to other cases where interacting non-Hermitian models serve as an effective description of dissipative dynamics~\cite{ Kazuki_2019,Chen_2023} and to understand the role of non-Markovianity in non-Hermitian thermalization~\cite{nh_sensing1,Mouloudakis_2022}. 

~\nocite{ETH_off_diagonal2,ETH_off_diagonal4,Gopalakrishan,forum,nh_MBL1,
nh_MBL2,Orito2023,book_breuer_petruccione,daley_2014,zoller_1,zoller_2,molmer_1993,
Lewenstein_2007,Dutta_2015,Gross_2017,Tarruell_2018,Takahashi_2020,chin_2010,
Trefzger_2011,Chomaz_2022,Gou_2020}

{\it Acknowledgements.---}
We gratefully acknowledge useful discussions with Philipp Uhrich.
We acknowledge support by the ERC Starting Grant StrEnQTh (project ID 804305), Provincia Autonoma di Trento, Q@TN, the joint lab between University of Trento, FBK-Fondazione Bruno Kessler, INFN-National Institute for Nuclear Physics and CNR-National Research Council, and support from ICSC - Centro Nazionale di Ricerca in HPC, Big Data and Quantum
Computing, funded by the European Union under NextGenerationEU. The project is funded within the QuantERA II Programme that has received funding from the European Union’s Horizon 2020 research and innovation programme under Grant Agreement No 101017733.
This project has received funding from the Italian Ministry of University and Research (MUR) through the FARE grant for the project DAVNE (Grant R20PEX7Y3A) and the European Union under Horizon Europe Programme - Grant Agreement 101080086 - NeQST. 
Funded by the European Union - Next Generation EU, Mission 4 Component 2 - CUP E53D23002240006.
S.S.R.\ acknowledges the faculty
research scheme at IIT (ISM) Dhanbad, India, under
Project No. FRS/2024/PHYSICS/MISC0110.
S.B.\ acknowledges CINECA for the use of HPC resources under Italian
SuperComputing Resource Allocation– ISCRA Class C
Projects No. ISSYK-2 (HP10CP8XXF) and No. DISYK
(HP10CGNZG9).

Views and opinions expressed are however those of the author(s) only and do not necessarily reflect those of the European Union or the European Commission.
Neither the European Union nor the granting authority can be held responsible for them.

\bibliography{bibliography}

\clearpage  
\onecolumngrid  

\begin{center}
\textbf{\large Supplemental Material: Unveiling Eigenstate Thermalization for Non-Hermitian systems}
\end{center}
\twocolumngrid 
\setcounter{section}{0}
\setcounter{equation}{0}
\setcounter{figure}{0}
\setcounter{table}{0}
\renewcommand{\thesection}{S\arabic{section}}
\renewcommand{\theequation}{S\arabic{equation}}
\renewcommand{\thefigure}{S\arabic{figure}}
\renewcommand{\thetable}{S\arabic{table}}

In this Supplemental Material (SM), we provide detailed elaborations on several discussions from the main text. We begin with the basic formalism of ETH for Hermitian systems in Sec. \ref{Appendix:A}. Section \ref{Appendix:B} then presents various cases of the evolution of initial states under non-Hermitian Hamiltonians. In Sec. \ref{Appendix:C}, we describe the numerical computations used to generate the plots presented both in the main text and in this SM. Here, we also extend our analysis to the other non-Hermitian variants of the SYK Hamiltonian. Section~ \ref{Appendix:D} explores ETH results for eigenstates in regions of the energy spectrum not covered in the main text. For completeness, we also include results obtained in the biorthogonal basis in Sec. \ref{Appendix:E}. Finally, Sec. \ref{Appendix:F} offers an in-depth discussion on the dynamical behavior of local observables under non-Hermitian evolution, benchmarking these dynamics against the static ETH results presented in the main text. We also address the experimental feasibility of one of the models considered, making our study experimentally realizable.

\section{ETH: Basic Formalism for Hermitian Systems}
\label{Appendix:A}
Though not proven, the ETH has been widely tested in Hermitian systems. Its central claim is that the expectation value of a generic local observable $\hat{O}$, when  computed in the energy eigenbasis of a non-integrable model, given by basis states $\{|m\rangle\}$, follows a surprisingly simple and general form \cite{ETH_ansatz7}: 
\begin{equation}
O_{mn}=\langle m|{\hat{O}}|n \rangle=O(\overline{E}) \delta_{mn}+e^{-S(\overline{E})/2} f(\overline{E},\omega) d_{mn}\,.
\label{eq:ETH_main}
\end{equation}
Here, $\overline{E}=\frac{E_m+E_n}{2}$ is the mean energy of the considered eigenstates, $\omega=|E_m-E_n|$ their absolute energy difference, 
$S(\bar{E})$ is the entropy, and $d_{mn}$'s are Gaussian random numbers (real or complex) with zero mean and unit variance  (see also \cite{ETH_off_diagonal1,ETH_off_diagonal2,ETH_off_diagonal3,ETH_off_diagonal4}). 
Important and non-trivial features are the smoothness of $O(\overline{E})$ and $f(\overline{E},\omega)$, as well as the strong decrease in fluctuations with density of states (entering through the entropy). 
As a result, different states with a given energy $E$ are---for what concerns local observables---indistinguishable from each other, as well as from a microcanonical ensemble at the same energy density. It is the purpose of this article to provide a translation of this concept to non-Hermitian systems and to show how it influence their out-of-equilibrium behavior.

\section{Time evolution under non-Hermitian Hamiltonian}
\label{Appendix:B}
\begin{figure}
 \includegraphics[width=\linewidth]{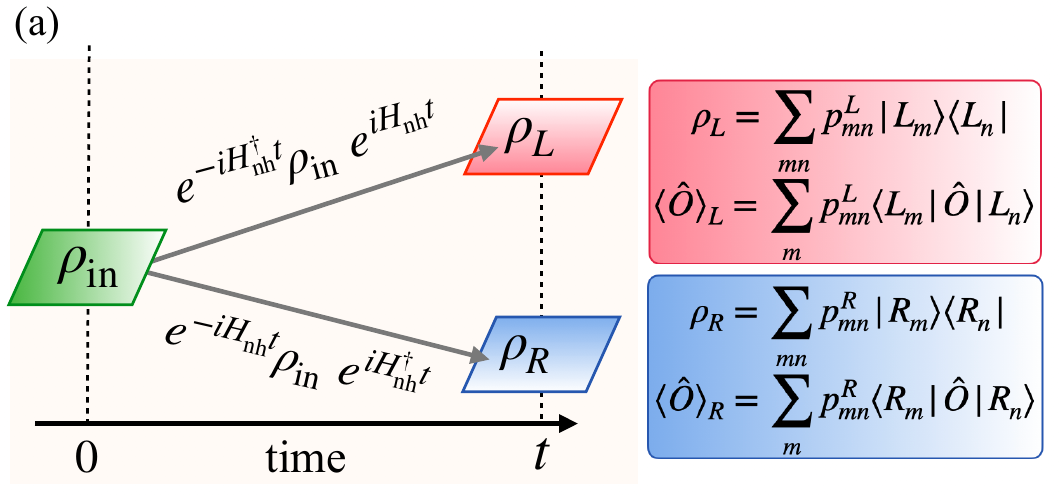}\\
 \caption{The evolution under a generic non-Hermitian Hamiltonian $H_{\mathrm{nh}}$ ($H_{\mathrm{nh}}^\dagger$) naturally leads to measuring the expectation value of an operator $\hat{O}$ in right (left) eigenvectors only.  }
 \label{sup:fig1}
 \end{figure}

To a considerable part, the significance of the ETH for Hermitian systems stems from its relevance to time evolution of closed quantum systems. In this section, we discuss the fate of an initial state when it is evolved under any generic non-Hermitian Hamiltonian $H_\mathrm{nh}$, as shown in the schematic presented in Fig.~\ref{sup:fig1}. 

For a non-Hermitian Hamiltonian $H_{\mathrm{nh}}$, an initial state $\rho_{\mathrm{in}}$ evolves into 
\begin{eqnarray}
\label{eqn:dynamics1}
\rho(t)&=&e^{-itH_{\mathrm{nh}}}\rho_{\mathrm{in}} e^{itH^\dagger_{\mathrm{nh}}}\\
&=&\sum_{mn} e^{-it\phi_{mn}} a_{mn} |R_m\rangle \langle R_n|,\nonumber
\end{eqnarray}
where, $a_{mn}=\langle L_m|\rho_\mathrm{in}|L_n\rangle$ and $\phi_{mn}=\epsilon_m-\epsilon_n^*$. 
Therefore, the expectation values of observables, $\Tr(\rho(t)\hat{O})$, will be determined by the matrix overlap elements $\langle R_n|\hat{O}|R_m\rangle $. In this regard, it is illustrative to first consider two  scenarios that can commonly arise, one that drives the system into a single eigenstate and another one that produces a diagonal ensemble in analogy to Hermitian time evolution, before coming to the general case.

\noindent 
{\bf Case I:}  {\it Complex $\epsilon_m$'s  with a non-degenerate $\max\{s_m\}=s_{\tilde{m}}$.} Separating diagonal and off-diagonal terms in Eq.~(\ref{eqn:dynamics1}), and introducing the decomposition of the complex energies in real and imaginary parts, $\epsilon_m=w_m+i s_m$, with $w_m, s_m \in \mathbb{R}$, we have  
\begin{align}
\rho(t)&=\sum_m e^{2t s_m}a_{mm}|R_m\rangle \langle R_m|\\
  & \qquad +\sum_{m\neq n}e^{-it(w_m-w_n)}e^{t(s_m+s_n)}a_{mn}|R_{m}\rangle \langle R_{n}|.\nonumber
\label{eqn:dynamics1}
\end{align}
At large times, $\rho(t)$ eventually converges to the right eigenvector  with slowest decay rate \cite{Gopalakrishan}, 
\begin{eqnarray}
\rho(t) \xrightarrow{t\rightarrow \infty}  e^{2t s_{\tilde {m}}}a_{\tilde{m}\tilde{m}}|R_{\tilde{m}}\rangle \langle R_{\tilde{m}}|.
\end{eqnarray}
Suitably normalized, observables will be given by $\langle R_{\tilde{m}}|\hat{O}|R_{\tilde{m}}\rangle$.

\noindent 
{\bf Case II:} {\it All  $\epsilon_m$'s real (Parity-Time symmetric case).} In this case, Eq.~(\ref{eqn:dynamics1}) reduces to  
\begin{equation}
\rho(t)=\sum_m a_{mm}|R_m\rangle \langle R_m|+\sum_{m\neq n} a_{mn} e^{-it(w_m-w_n)}|R_{m}\rangle \langle R_{n}|\,.
\end{equation}
As in the Hermitian case assuming irrelevance of resonances at large times, this state evolves into a diagonal ensemble consisting of the right eigenvectors \cite{Gopalakrishan,time_evol_righteigen}, 
\begin{equation}
\rho(t) \xrightarrow{t\rightarrow \infty} \sum_m a_{mm} |R_{m}\rangle \langle R_{m}|.
\end{equation}
Also here, observables at long times are determined by (suitably weighted) matrix elements $\langle R_{m}|\hat{O}|R_{n}\rangle$. 

\noindent 
{\bf General Case:} \emph{Arbitrary non-Hermitian time evolution.}
For an arbitrary non-Hermitian time evolution of any observable $\hat{O}$, Eq.~\eqref{eqn:dynamics1} predicts
\begin{eqnarray}
\text{Tr}(\hat{O}\rho(t))&=&\sum_{mn} e^{-it\phi_{mn}} a_{mn} \text{Tr}(\hat{O}|R_m\rangle \langle R_n|),\nonumber\\
&=&\sum_{mn} e^{-it\phi_{mn}} a_{mn} \sum_k \langle R_k|\hat{O}|R_m\rangle \langle R_n|L_k\rangle,\nonumber\\&=&\sum_{mn} e^{-it\phi_{mn}} a_{mn}  \langle R_n|\hat{O}|R_m\rangle\,.
\end{eqnarray}
Here, we have taken the trace in the biorthogonal basis that satisfies the completeness relation $\sum_m |R_m\rangle \langle L_m|=\mathbb{I}$, and used the relation $ \langle R_n|L_k\rangle=\delta_{nk}$. 
Again, observables are evaluated as a (time-evolving) function of matrix elements in the right eigenvectors. 

As these considerations illustrate, states determined by the right eigenvectors of the model are naturally prepared. This operational interpretation justifies the preferability of the associated basis over the biorthogonal basis when we aim at the computation of physically relevant quantities. Similar justification also holds for the set of left eigenvectors $\{\langle L_m|\}$ of the model,  which can be realized when we consider the evolution of an initial state under $H_{\mathrm{nh}}^\dagger$. A schematic presented in Fig. \ref{sup:fig1} shows how the evolution under a generic non-Hermitian Hamiltonian $H_{\mathrm{nh}}$ ($H_{\mathrm{nh}}^\dagger$) naturally leads to measuring the expectation value of an operator $\hat{O}$ in right (left) eigenvectors only.

\section{Details on numerical computations}
\label{Appendix:C}
In this section, we elaborate on the numerical techniques we have employed to obtain the main results of our work. 
We start our discussion by describing the construction of the random non-Hermitian matrix $H_{\mathrm{nh}}$ using the Ginibre ensemble. Subsequently, we detail the construction and diagonalization of the non-Hermitian Sachdev--Ye--Kitaev model. Furthermore, we compare the results for other choices of non-Hermitian SYK Hamiltonians.

\subsection{Non-Hermitian random matrices}

Given any non-Hermitian Hamiltonian $H_{\mathrm{nh}}$, we can obtain its set of left and right eigenvectors using the  similarity transformation given by
\begin{eqnarray}
H_{\mathrm{nh}}=P D P^{-1}=\sum_m \epsilon_m |R_m\rangle \langle L_m|,
\label{eqn:eigen_decomposition}
\end{eqnarray}
where the vectors $|R_m\rangle$ ($\langle L_m|$)  construct the columns (rows) of $P$ ($P^{-1}$) and  $D$ is a diagonal matrix with entries $D_{mm}=\epsilon_m$. As a result, we have $\langle L_m| R_n \rangle =\delta_{mn}$. In addition, we choose the normalization $\langle R_m| R_m \rangle =1$. 

For the  random ensemble study, we construct the non-Hermitian random matrix $H_{\mathrm{nh}}$ by drawing its elements from the complex Ginibre ensemble. We start with a representation of  $H_{\mathrm{nh}}$ in the fermionic basis  as follows
\begin{equation}
    H_{\mathrm{nh}}=\sum_{mn}c^\dagger_m \mathcal{H}_{mn} c_n.
    \label{eq:Hnh_fermionic-basis}
\end{equation}
We next choose $\mathcal{H}_{mn}=\mathcal{A}_{mn}+i\,  \mathcal{B}_{mn}$, 
where $\mathcal{A}_{mn}$ and $\mathcal{B}_{mn}$ are chosen independently from a Gaussian distribution with zero mean  and variance $\frac{1}{2\mathcal{D}}$. We then use the exact diagonalization technique to bring the Hamiltonian $\mathcal{H}$ in the form as given in Eq.~(\ref{eqn:eigen_decomposition}). 

Using the representation given in Eq.~\eqref{eq:Hnh_fermionic-basis}, it is also straightforward to define fermionic local observables that draw close analogy to the investigation on the non-Hermitian SYK model. As stated in the main text, for our purposes we choose $\hat{O}=\hat{n}_i=c_i^\dagger \hat{c}_i$, where due to symmetry the mode $i$ can be chosen arbitrarily. 

\begin{figure}
\includegraphics[width=8.2cm]{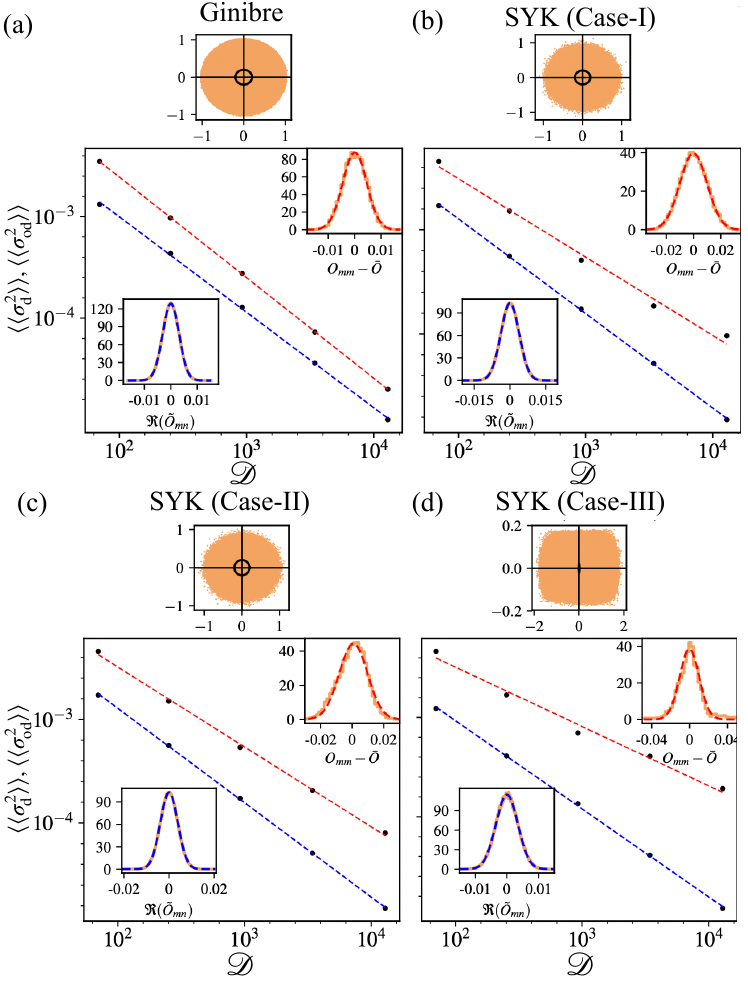}\\
  \caption{
  Generic behavior of fluctuations of local operators computed for the eigenbasis of SYK model for different choices of complex interaction elements $J_{i_1i_2:j_1j_2}$. In all the cases, we plot the decay of fluctuations of diagonal ($\langle \langle \sigma_{\mathrm{d}}^2\rangle \rangle$, linearly fitted with dotted red line) and off-diagonal matrix elements ($\langle \langle \sigma_{\mathrm{od}}^2\rangle \rangle$, linearly fitted with dotted blue line) with the Hilbert space dimension $\mathcal{D}$ (for $N=8, 10, 12, 14, 16$) and show the Gaussian fits of the data for $N=16$ in the insets. Additionally, on top of each plot, we mark the region of the energy subspace from which the eigenstates are sampled. For comparison,  we also present the results already shown in the main text: (a) the Ginibre ensemble and (b) the non-Hermitian SYK model Case I. The qualitative feature remain the same for different choices of the complex interaction elements $J_{i_1i_2:j_1j_2}$, (c) SYK Case II with  real random diagonal elements but complex random off-diagonal elements satisfying $J_{i_1i_2:j_1j_2}^* \neq J_{j_1j_2:i_1i_2}$; (d)  SYK Case III, with complex random diagonal elements and complex random off-diagonal elements with $J_{i_1i_2:j_1j_2}^* = J_{j_1j_2:i_1i_2}$. All the plots clearly show linear decay of fluctuations with $\mathcal{D}$  in the log--log scale. The qualitative agreement of the plots thus shows the generality of the ETH features, independent of the choice of complex couplings. }
\label{fig:Compare_SYK}
\end{figure}

\subsection{Non-Hermitian Sachdev--Ye--Kitaev model}
The interaction amplitudes in the standard Hermitian SYK Hamiltonian of complex fermions $H^{\text{SYK}_4}$ are considered as zero mean Gaussian random variables with 
$\text{var}(\text{Re}[J_{i_1 i_2;j_1 j_2}])= J^2, \text{var}(\text{Imag} [J_{i_1 i_2;j_1j_2}]) = 0$ if $i_1 = j_1, i_2 = j_2$, and  $\text{var}(\text{Re}[J_{i_1 i+2; j_1j_2}]) = \text{var}(\text{Im}[J_{i_1 i_2;j_1 j_2}]) =J^2/2$, otherwise. Usually, the overall scale $J$ is set to unity.

In case of the non-Hermitian SYK Hamiltonian, we generically choose  all the amplitudes to be complex random variables, i.e., $J_{i_1i_2:j_1j_2}=\alpha_{i_1i_2:j_1j_2}+ i \beta_{i_1i_2:j_1j_2}$, where $\alpha_{i_1i_2:j_1j_2}$ and $\beta_{i_1i_2:j_1j_2}$ are independent random numbers chosen from Gaussian distribution with zero mean and variance $1/2$. We additionally consider $J_{i_1i_2:j_1j_2}^* \neq J_{j_1j_2:i_1i_2}$, for the off-diagonal elements.  

Similar to the Hamiltonian matrix construction for the Hermitian SYK model as mentioned in Ref.~\cite{SYK5}M, we construct the non-Hermitian Hamiltonian matrix in the fermionic Fock basis states $|s_a\rangle$ for the half-filled sector, so $a = 1,\ldots,\mathcal{D}$. The matrix elements are then computed as $\langle s_a|H^{\text{SYK}_4}_{\mathrm{nh}}| s_b \rangle $, where $s_a$ and $s_b$ represent $N$-bit strings of equal number of $0$ and $1$ bits denoting empty and occupied fermionic modes, respectively. For the Hamiltonian in Eq.~\eqref{eqn:SYK},
the non-zero matrix elements correspond to the Hamming distance $d(s_a,s_b) = 0, 2, 4$. We choose these elements as complex random variables with properties mentioned in the previous paragraph. While populating the elements into the Hamiltonian matrix, we ensure the antisymmetry of the interaction amplitudes under the  permutation of the subscripts. In this way, a single realization of the non-Hermitian SYK Hamiltonian is constructed. We diagonalize this Hamiltonian in the form of Eq.~(\ref{eqn:eigen_decomposition}) in order to compute the relevant properties.

For both the Ginibre ensemble and the non-Hermitian SYK Hamiltonian, we evaluate diagonal and off-diagonal matrix elements of observables with respect to the eigenstates of the diagonalized Hamiltonian. In addition, we consider multiple realizations  to improve the statistics in our analysis. Since we perform full-diagonalization of the matrices, we employ \verb+MPI+ \cite{forum} methods to solve different realizations in parallel, which reduces the computation time significantly.

\subsection{Other choices of Non-Hermitian SYK model}
\label{syk_incar}

In this subsection, we show that our results generically apply for other choices of non-Hermiticity of the SYK Hamiltonian. All the results presented in the main text are obtained for the choice mentioned in the previous subsection, which we refer to as Case I. However, we can also consider real random diagonal elements, but complex random off-diagonal elements satisfying $J_{i_1i_2:j_1j_2}^* \neq J_{j_1j_2:i_1i_2}$. We call this choice Case II. Furthermore, we can choose the diagonal elements as complex random variables, but with $J_{i_1i_2:j_1j_2}^* = J_{j_1j_2:i_1i_2}$, which we refer to as Case III.  

In Fig.~\ref{fig:Compare_SYK}, we illustrate and compare the key characteristics of the ETH for these different choices of the non-Hermitian SYK Hamiltonian. For comparison, we also reproduce the results of the Ginibre ensemble and SYK Case I in (a) and (b), which have already been presented in the main text. The overlaying complex eigenvalues of multiple realizations are shown at the top of each panel corresponding to the different cases. These also show the region of the eigenstates selected for our analysis. Like in the main text, we obtain exponentially suppressed and 
Gaussian fluctuations for the diagonal and off-diagonal elements of the observable. The exponential law with respect to system-size is demonstrated by the linear trend of 
$\langle \langle \sigma_{\mathrm{d}}^2 \rangle \rangle$ and $\langle \langle \sigma_{\mathrm{od}}^2 \rangle \rangle$ with respect to $\mathcal{D}$ in the $\mathrm{log}$-$\mathrm{log}$ scales.
Therefore, our conclusions do not depend on the microscopic details of the considered non-Hermitian SYK model.

\section{ETH for eigenstates within slices of complex energy space of random \texorpdfstring{$H_{\mathrm{nh}}$}{Hnh}}
\label{Appendix:D}
\begin{figure}[ht]
 \includegraphics[width=1\linewidth]{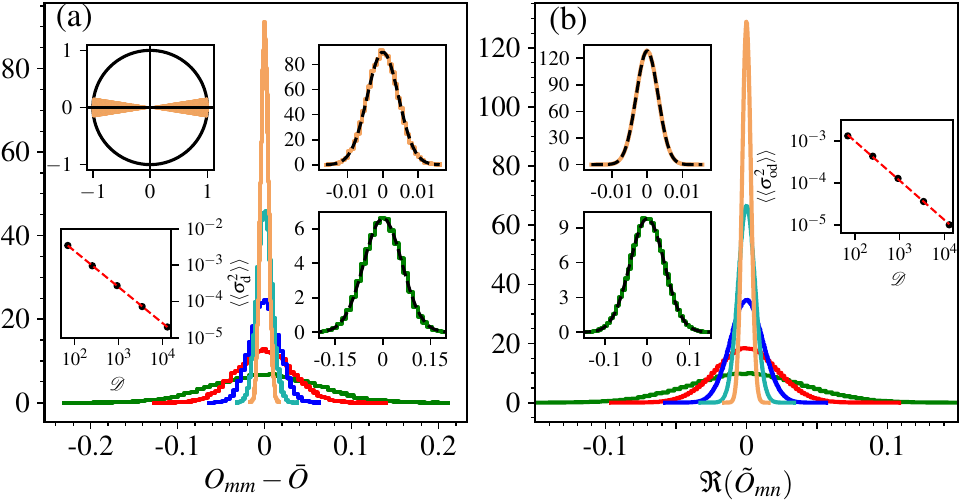}
  \caption{Signature of ETH in a random non-Hermitian Hamiltonian from the Ginibre ensemble. We now consider eigenstates with energy $\epsilon_m=|\epsilon_m|e^{i \Phi}$  lying within slices parameterized by $0 \leq|\epsilon_m|\leq 1$ and $0 \leq |\Phi| \leq 0.05$ of the complex energy spectrum of the model. 
  (a) Histogram of the diagonal terms $O_{mm}$ obtained for system sizes and number of random realizations as in Fig.~\ref{fig:ROR_corrected_Ginibre} of the main text. 
  The right insets show the Gaussian fittings (black dashed) of the histograms for $N=8$ (bottom, 5000 random realizations) and $N=16$ (top, 20 random realizations), respectively. 
  Bottom left: the variance of the diagonal data $\langle \langle \sigma_{\mathrm{d}}^2 \rangle \rangle$ decays polynomially with Hilbert space dimension $\mathcal{D}$. 
  (b) Similar analysis for the modified off-diagonal terms $[\widetilde{O}_{mn}]_{m\neq n}$.   All the behavior obtained here strongly mimics the one obtained in Fig.~\ref{fig:ROR_corrected_Ginibre} of the main text, justifying the generality of our formalism when applied to a set of states chosen from any part of the complex eigenspectrum of the model.}
  \label{fig:angsamp}
\end{figure}

In this section, we further substantiate our numerical study done in the main text and shown in Fig.~\ref{fig:ROR_corrected_Ginibre}. 
That figure unveils the ETH signature for the random non-Hermitian model constructed using the Ginibre ensemble for the eigenstates lying within the disk of radius $0 \leq |\epsilon_m| \leq 0.2$. 
Here, we perform a similar analysis but for eigenstates with energies $\epsilon_m=|\epsilon_m|e^{i\Phi}$ such that $0 \leq|\epsilon_m|\leq 1$ and $\Phi$ lies within $0.0\leq |\Phi| \leq 0.05$. 
In Fig.~\ref{fig:angsamp} of this SM, we plot the behavior for the diagonal terms $O_{mm}$ (panel a) and for the modified off-diagonal terms $[\widetilde{O}_{mn}]_{m \neq n}$ (panel b). All the other parameters are the same as considered in Fig.~\ref{fig:ROR_corrected_Ginibre}. The behavior exactly mimics the one that has been obtained in Fig.~\ref{fig:ROR_corrected_Ginibre} and shows the generality of our formalism when applied to eigenstates chosen from any portion of the complex energy spectrum of the model. 

\section{Results using biorthogonal basis}
\label{Appendix:E}
\begin{figure}[ht]
 \includegraphics[width=1\linewidth]{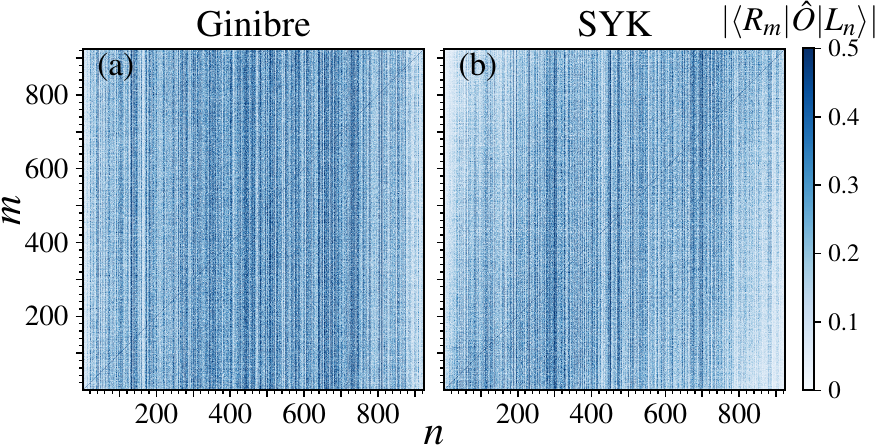}
  \caption{Expectation value of a local operator, computed in the biorthogonal basis of non-Hermitian Hamiltonians for a single random realization. 
  (a) Results for a random matrix $H_{\mathrm{nh}}$ obtained from the Ginibre ensemble and (b) for the non-Hermitian SYK model for $N=12$. In both cases, $|\langle R_m|\hat{O}|L_n\rangle|$ deviates significantly from the ETH prediction as the plots exhibit structured patterns, suggesting strong off-diagonal fluctuations that are comparable to the diagonal elements. This behavior is characteristically different from that obtained for right (or left) eigenvectors of the models, as shown in Fig.~\ref{fig:ROR_Ginibre} in the main text.}
  \label{fig:ROL}
\end{figure}

In this section, for completeness we present a summary of the results obtained using the biorthogonal basis, for both random matrices $H_{\mathrm{nh}}$ using the Ginibre ensemble as well as the non-Hermitian SYK model as described in Eq.~(\ref{eqn:SYK}) of the main text. We consider the same local observable $\hat{O}$ as in the main text. Figure~\ref{fig:ROL} depicts the behavior of $|\langle R_m|\hat{O}|L_n\rangle|$ obtained for a single random realization with $N=12$. For both models, the behavior of $|\langle R_m|\hat{O}|L_n\rangle|$ remains far from the ETH prediction. In particular, the plots exhibit a ``structured" pattern for off-diagonal elements, which signatures off-diagonal and diagonal elements being of the same order. Similar behavior is also reported in Ref.~\cite{ETH_nh}. 

This behavior is in stark contrast with the elements evaluated in the right (or left) eigenvectors, which, as discussed in the main text, do exhibit behavior compatible with ETH. 
In contrast to $\langle R_m|$ and $|R_n\rangle$, $\langle R_m|$ and $|L_n\rangle$ are already orthogonal, so a modification to bring predictions into accordance with ETH analogously to Eq.~\eqref{eq:ETH_modified_substraction} of the main text does not seem straightforward.

\section{Certification of non-Hermitian ETH by non-Hermitian dynamics}
\label{Appendix:F}
In this section, we discuss the thermalization dynamics of quantum many-body systems subjected to non-Hermitian evolution given by
\begin{equation}
|\psi(t)\rangle = \frac{e^{-iH_{\rm nh}t}|\psi(0)\rangle}{||e^{-iH_{\rm nh}t}|\psi(0)\rangle||},
\label{eq:renom_evo}
\end{equation}
where we normalized the state to preserve probabilities.  Then, the evolution of a local observable is computed as $O(t)=\langle\psi(t)|\hat{O}|\psi(t)\rangle$.

We match the long-time characteristics of local operators with the predictions made by the non-Hermitian ETH framework presented in the main text. To begin with, we present the dynamics of the Ginibre and non-Hermitian incarnations of the SYK model, which are our main test beds for the non-Hermitian ETH framework. Subsequently, we extend our analysis to an experimentally feasible non-Hermitian many-body model, the interacting disordered Hatano--Nelson model, and examine its thermalization dynamics in the ergodic regime. Furthermore, we illustrate the evolution in the many-body localized  regime, where the model is not expected to thermalize. These results further substantiate the non-Hermitian ETH framework in understanding the thermalization properties of a non-Hermitian evolution. 

\subsection{Framework}
Let us recall the evolution of any observable $\hat{O}$ under an arbitrary non-Hermitian Hamiltonian, given by
\begin{equation}
\langle \hat{O} \rangle_{\rho}=\sum_{mn} e^{-it\phi_{mn}} a_{mn}  \frac{\langle R_n|\hat{O}|R_m\rangle}{\text{Tr}[\rho]}\,.
\end{equation}
Now, applying the ETH predictions as formulated in Eqs.~({\color{red}7})-({\color{red}8}) of the main text, we obtain
\begin{widetext}
\begin{eqnarray}
\langle \hat{O} \rangle_{\rho}&=& \sum_{m=n}  e^{-it\phi_{mm}} a_{mm}  \frac{ \Big(\overline{O}+ Q_{mm} \sqrt{\frac{\overline{O^2}}{\mathcal{D}}} \Big)}{\text{Tr}[\rho]}+\sum_{m\neq n} e^{-it\phi_{mn}} a_{mn}  \frac{ \Big(\overline{O} \langle R_m|R_n\rangle + S_{mn} \sqrt{\frac{\overline{O^2}}{\mathcal{D}}} \Big)}{\text{Tr}[\rho]}, \nonumber\\
&=& \sum_{mn} e^{-it\phi_{mn}} a_{mn}  \frac{ \Big(\overline{O} \langle R_m|R_n\rangle \Big)}{\text{Tr}[\rho]}+ \sum_{m n} e^{-it\phi_{mn}} a_{mn}  \frac{\sqrt{\frac{\overline{O^2}}{\mathcal{D}}} \ell_{mn}}{\text{Tr}[\rho]}, \nonumber
\end{eqnarray}
where  $\ell_{mm}=Q_{mm}$, $\ell_{mn}=S_{mn}$. 
Using the relation $\text{Tr}[\rho]=\sum_{mn} e^{-it\phi_{mn}} a_{mn}   \langle R_m|R_n\rangle$, we get
\begin{eqnarray}
\langle \hat{O} \rangle_{\rho}&=& \overline{O}+  \sqrt{\frac{\overline{O^2}}{\mathcal{D}}} \frac{\sum_{m n} e^{-it\phi_{mn}} a_{mn}  \ell_{mn}}{\sum_{mn} e^{-it\phi_{mn}} a_{mn}   \langle R_m|R_n\rangle}.
\label{eqn:dynamics}
\end{eqnarray}
\end{widetext}
In the above expression, the second term becomes insignificant for large system sizes as well as after performing time averaging. Then, the expectation value of the observable converges to the mean of the observable, $\overline{O}$ at long time. Therefore, the formalism clearly demonstrates that if the ETH ansatz holds for a non-Hermitian Hamiltonian, its imprint will be evident in the long-time dynamics of the expectation values of local observables. In particular, under the evolution of non-Hermitian random matrix like models, a local observable will thermalize to its mean value.

In the following, we discuss the dynamical characteristics of local observables when different initial states undergo non-Hermitian evolution, and
demonstrate their consistency with the Eq.~(\ref{eqn:dynamics}).

\begin{figure}[ht]
 \includegraphics[width=1\linewidth]{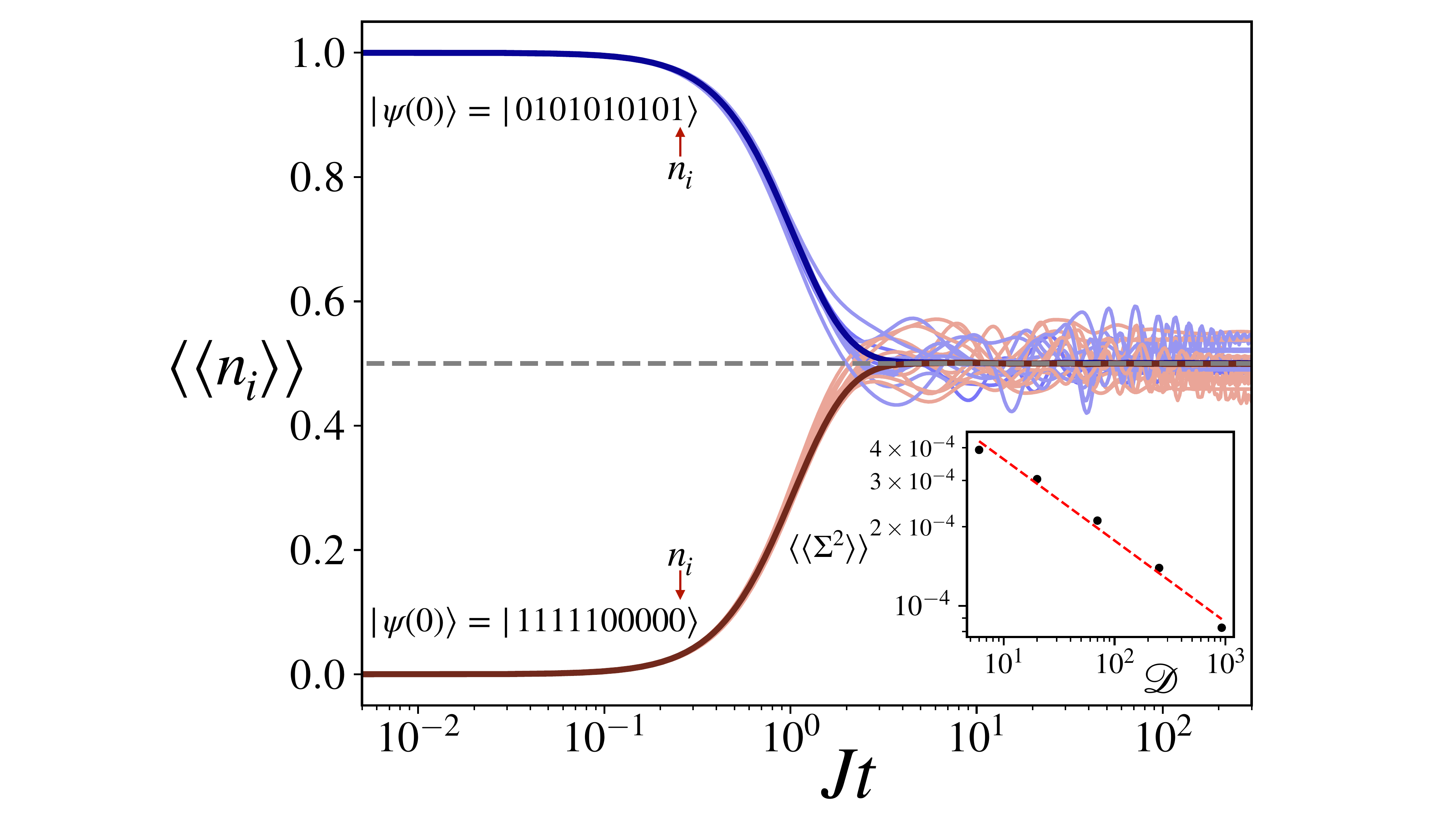}
  \caption{Thermalization dynamics of local number operator under a non-Hermitian Hamiltonian.  
  Results are shown for the random matrix $H_{\mathrm{nh}}$ obtained from the Ginibre ensemble, for $N=10$ at the half filling sector, for two initial states $|\psi\rangle=|0101\dots01\rangle$ (blue) and $|\psi\rangle=|1\rangle^{\otimes^{N/2}} \otimes |0\rangle^{\otimes^{N/2}}$ (red). The light shaded curves show the dynamics for $10$ different realizations of the Hamiltonian, whereas the dark curves represent the corresponding ensemble averaged dynamics over $3000$ samples of the Ginibre ensemble. 
  At long times, the local number operator fluctuates around the operator mean-value $\overline{n}_{i}=0.5$ (dashed grey line) yielding a long-time average close to this value. The ensemble averaged evolution, $\langle\langle n_{i}\rangle \rangle$, converges to this mean value as illustrated by the dark blue and red curves. Inset: Ensemble averaged long-time fluctuation plotted against the Hilbert-space dimension. The fluctuations get suppressed with the increase of system size. This strongly suggests that the ETH prediction derived and tested for the Ginibre ensemble is reflected in the long-time dynamics of local observable. }
  \label{fig:dyn_gin}
\end{figure}

\subsection{Evolution under Ginibre ensembles}
To illustrate the dynamics under the Hamiltonian constructed by choosing the elements from a Ginibre ensemble, we start with two different initial states, 
$|\psi(0)\rangle=|0101\dots 01\rangle$, and $|\psi(0)\rangle=|1\rangle^{\otimes^{N/2}} \otimes |0\rangle^{\otimes^{N/2}}$. These states are then evolved as in Eq.~\eqref{eq:renom_evo}, after which we compute the time-evolved expectation value of the number operator $\hat{n}_i=\hat{c}_i^\dagger \hat{c}_i$ of the $i$-th site.
In Fig.~\ref{fig:dyn_gin}, we depict the the ensemble averaged evolution as well as the dynamics of a few typical realizations of the Ginibre Hamiltonian. The dynamics displays a clear  thermalization process.
For the considered initial states,  
$n_i$ starts with its maximum or minimum values, 0 and 1, respectively. 
At long time, curves corresponding to individual realizations  fluctuate around the mean value (alternatively, infinite temperature average) of the operator $\overline{n}_{i}=0.5$. Consequently, the long-time average of $n_{i}(t)$ converges to $\approx 0.5$, as predicted by the ETH. The ensemble averaged curves, which are expected to capture large system size behaviour, show this convergence with even better clarity. 
Already for finite-size simulations, the fluctuations diminish with increasing system size. To illustrate this further, we compute the long-time variance of the fluctuation $\Sigma^2, $ and examine its ensemble-averaged behaviour with respect to the increase of Hilbert space dimension. The inset in Fig.~\ref{fig:dyn_gin} shows the ensemble-averaged long-time fluctuation with respect to the Hilbert space dimension, and demonstrates that the fluctuations get suppressed with increasing system size. 
These findings are consistent with the long-time behaviour predicted in Eq.~(\ref{eqn:dynamics}) and strongly align with the non-Hermitian ETH formalism as detailed in the main text.

\begin{figure*}[ht] 
\includegraphics[scale=0.265]{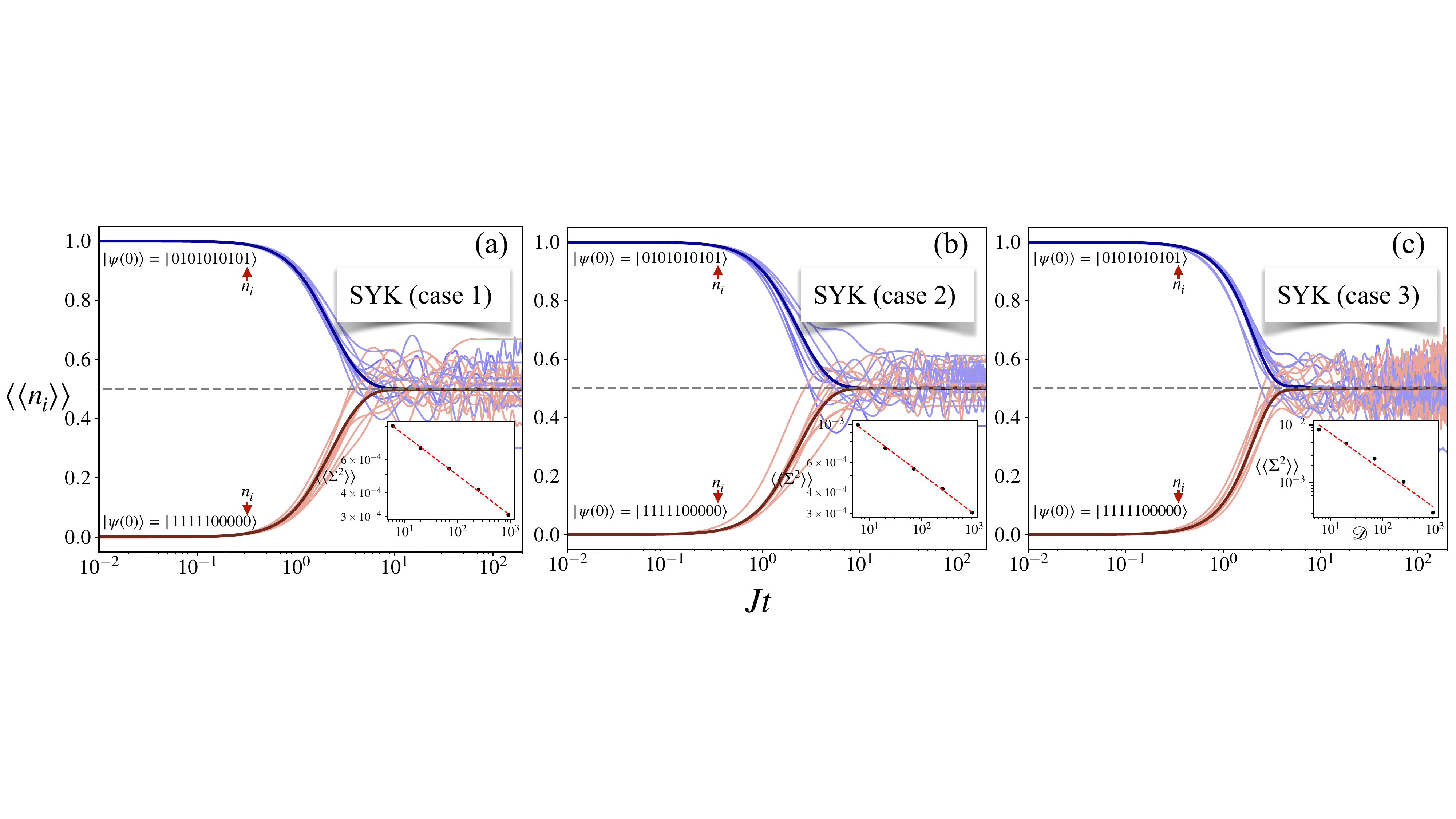}
\vspace{-3cm}
\caption{Thermalization dynamics of local number operator under the non-Hermitian variants of the SYK model considered in this work (see Sec.~\ref{syk_incar}). Data for $N = 10$ at half filling. The curves with light shading correspond to dynamics under typical realizations of the model. The darker curves represent the ensemble averaged dynamics over $4500$ samples. Individual evolutions oscillate around the average value. As a consequence, both temporal and ensemble average quickly converge. Insets: the fluctuations around the long-time averages decrease with the Hilbert-space dimension of the system. These findings are perfectly aligned with the non-Hermitian ETH predictions.}
  \label{fig:dyn_sykcases}
\end{figure*}

\subsection{Evolution under non-Hermitian SYK Hamiltonian}
For the same initial states, we now examine the dynamics under the non-Hermitian incarnations of the SYK model. Figure~\ref{fig:dyn_sykcases}(a)-(c) depicts the evolution for the Case I (as in main text), as well as Case II and Case III (see Sec.~\ref{syk_incar}). 
Similar to the Ginibre ensemble, the evolution of $n_i$ converges to the observable mean value $\overline{n}_i=0.5$ at long times. Consequently, the evolution results in locally thermalizing the system to infinite temperature. Convergence to the infinite-temperature averages of local observables has also been noted in the out-of-equilibrium dynamics of the Hermitian SYK model~\cite{SYK5}. In the present case, the dynamics under the individual realizations show relatively larger  fluctuations around the mean-value compared to the Ginibre case. However, as before, the fluctuations get suppressed with increasing system size. Therefore, the dynamics under the non-Hermitian SYK Hamiltonians show thermalization characteristics well aligned with the non-Hermitian ETH characteristics presented in the manuscript.

\subsection{Evolution under Hatano--Nelson model}
Furthermore, we demonstrate the thermalization dynamics of a non-Hermitian model of which quantum simulation is amenable to modern day experimental set-ups. In particular, we consider the interacting and disordered version of the Hatano--Nelson (HN) model~\cite{Hatano1, Hatano2, Hatano3}. The Hamiltonian of the model is given by 
\begin{equation}
    H_{\mathrm{HN}}=\sum_{i=1}^{N} -J( e^\chi \hat{c}^\dagger_i \hat{c}_{i+1}+e^{-\chi} \hat{c}^\dagger_{i+1} \hat{c}_{i})+V \hat{n}_i \hat{n}_{i+1} + h_{i}\hat{n}_i\,.
\label{eq:ham_hnmodel}
\end{equation}
This model describes a system of itinerant spinless fermions or hardcore bosons in a lattice.
We consider the model with periodic boundary conditions. The nearest-neighbour hopping strength $J=1$ scales the Hamiltonian, $\chi$ introduces non-reciprocity in the hopping process, and $V$ is the strength of nearest-neighbour interactions. The model is diagonally disordered by the onsite potentials $h_{i}$, which are independent random numbers chosen from the uniform distribution $[-W, W]$, where $W$ controls the strength of the disorder. This model is known to exhibit many-body localization for strong disorder~\cite{nh_MBL1,nh_MBL2,Orito2023}, while it is in the ergodic regime for small $W$. For all the simulations presented in this work, we consider $\chi = 0.1$ and $V=2$.


\subsubsection*{ETH analysis}

\begin{figure}[ht]
 \includegraphics[width=1\linewidth]{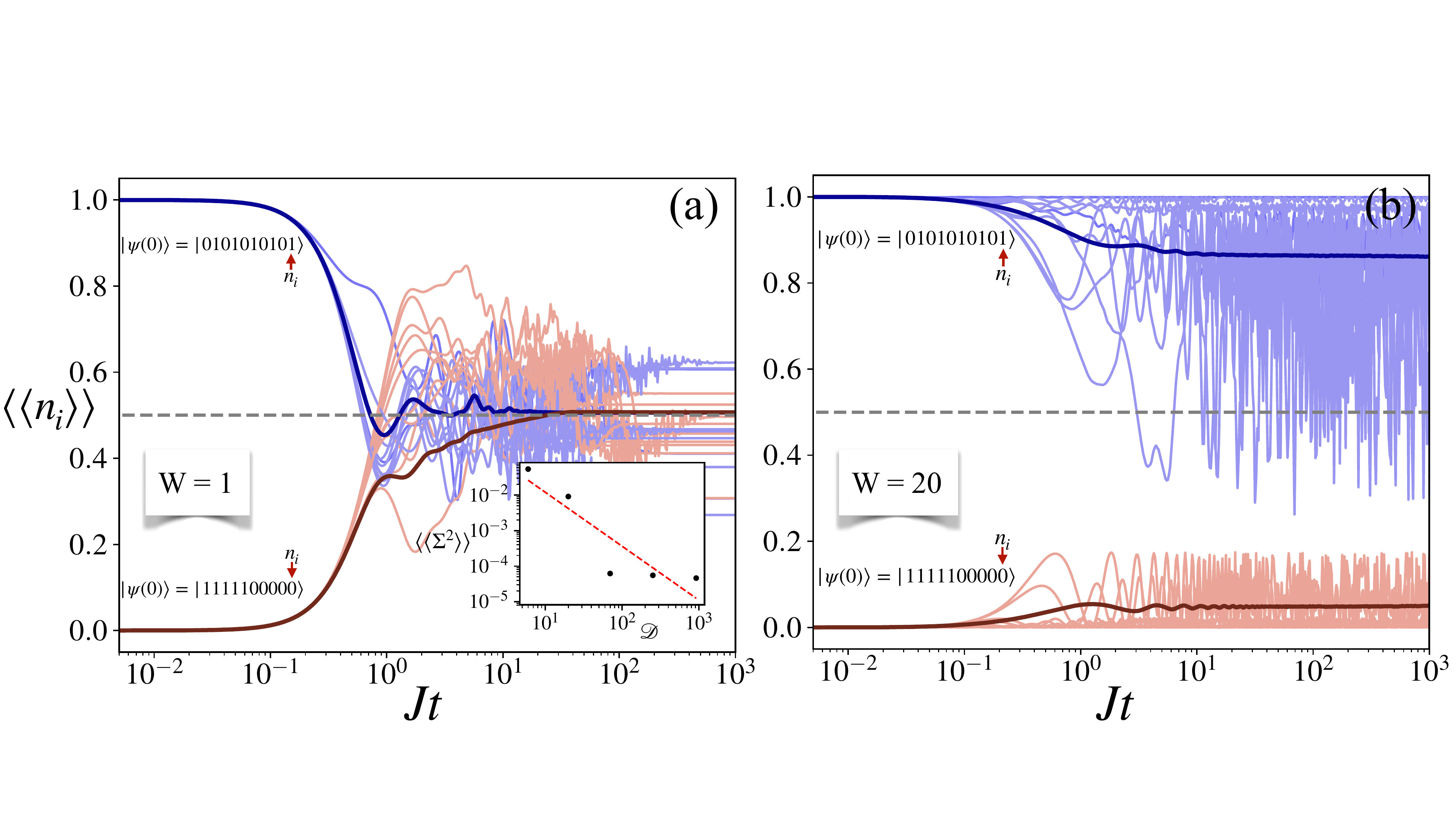}
  \caption{Dynamics of the number operator $\hat{n}_i$ computed  for an initial state evolving under the interacting and disordered Hatano--Nelson model. The initial states considered are the same as those in Fig.~\ref{fig:dyn_gin}. Curves with light shading show the dynamics for individual disorder realizations and the darker curves portray the ensemble-averaged dynamics over $3800$ samples. (a) Behavior for low disorder strength ($W=1$), where the system is in the ergodic regime. Here, for both the initial states $n_i$ tends to saturate to the value for an infinite temperature state, $\overline{n}_i=0.5$, suggesting thermalization in the system. The inset shows the fluctuations get suppressed with the increasing system size. The non-Hermitian ETH is expected to hold in this regime.
  In contrast, panel (b) illustrates the behavior for strong disorder ($W=20$). In this case, the system remains far from thermalization, and the number operator indicates a strong localization effect.}
  \label{fig:hatnel_numop}
\end{figure}

Similar to the previously discussed random non-Hermitian models, we examine the dynamics of the local number operator under the HN model for two different disorder strengths, $W=1$ and $W=20$. In this context, the model Hamiltonian is expected to be in the ergodic and localization regimes, respectively. The long-time dynamics of  $n_i$ should reflect these characteristics. In Fig.~\ref{fig:hatnel_numop} we depict the evolution for the previously considered initial states.  For the lower disorder strength $W=1$, the behavior of $n_i$
  remains similar to previous cases. On average the long-time evolution saturates close to the mean value of $n_i(t)$, and this is independent of initial states. Due to the physical structure of the model, the long-time fluctuations are more pronounced than the Ginibre ensemble and SYK models. However, with the increase of system size the fluctuations decrease. In contrast, for the larger disorder strength $W=20$, $n_i (t)$ remains close to its initial value even at long times, indicating that the memory of the initial state strongly persists. This behavior signals the presence of significant localization effects and corresponds to a regime where non-Hermitian ETH does not hold.

\begin{figure}[ht]
 \includegraphics[width=1\linewidth]{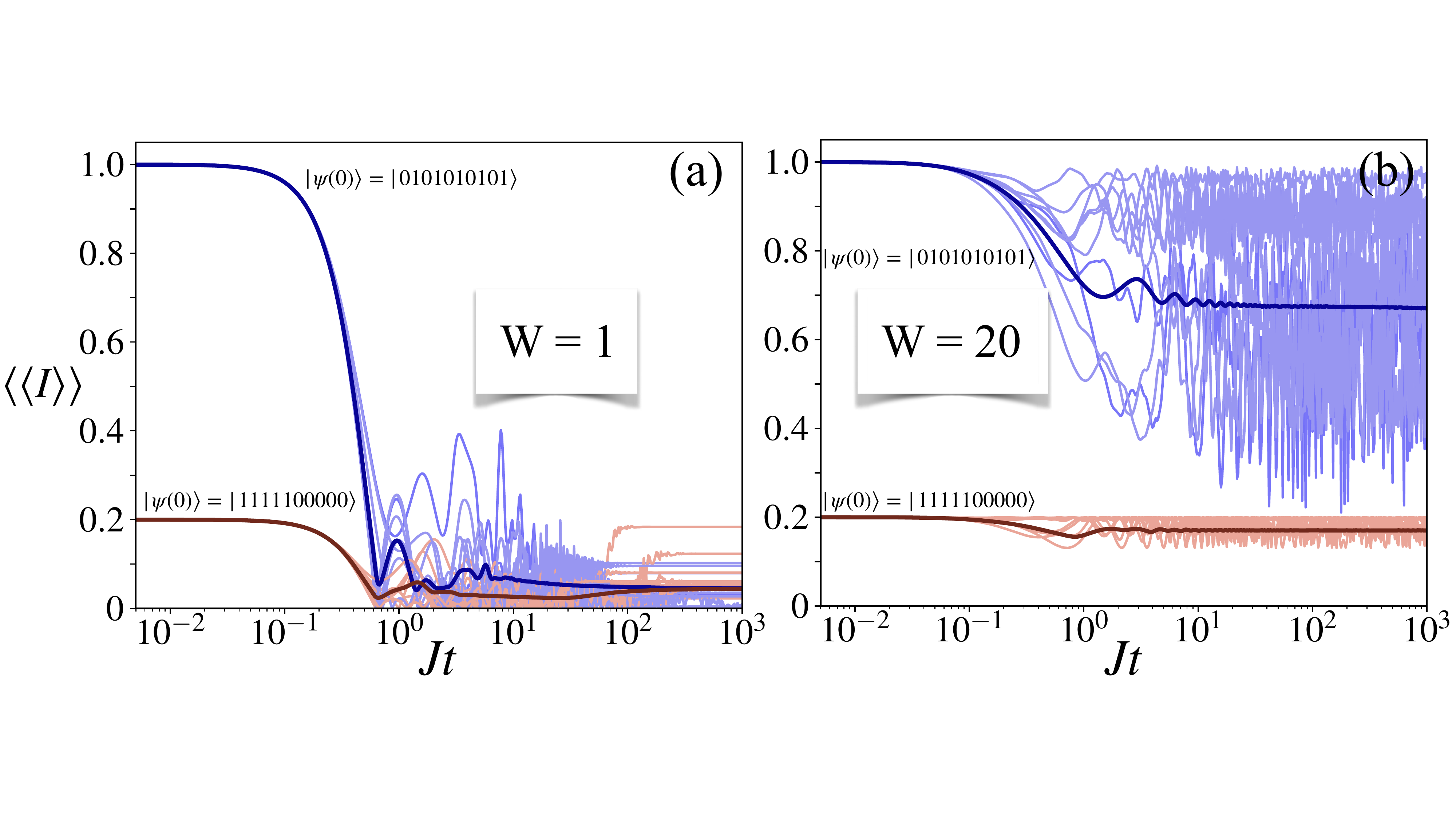}
  \caption{Dynamics of the particle imbalance parameter $I(t)$  computed  for an initial state evolving under the non-Hermitian Hatano--Nelson model. The initial states considered are the same as those in Fig.~\ref{fig:dyn_gin}. Panel (a) shows the behavior for low disorder strength ($W=1$), where the system is in the ergodic regime. Here, for both the initial states $|\psi\rangle=|0101\dots 01\rangle$ (red) and $|\psi\rangle=|1\rangle^{\otimes^{N/2}} \otimes |0\rangle^{\otimes^{N/2}}$  (blue) the imbalance parameter decays rapidly and saturates close to zero, suggesting thermalization. In contrast, panel (b) illustrates the behavior for strong disorder ($W=20$). In this case, the system remains far from thermalization, and the imbalance parameter indicates a strong localization effect.}
  \label{fig:imbal}
\end{figure}

As a second measure, we consider another experimentally feasible direct probe of thermalization, the particle imbalance
defined as 
\begin{eqnarray}
I(t) = \frac{2}{N}\Big|\sum_{i=1}^{N}(-1)^i \langle \psi(t)|\hat{n}_i|\psi(t)\rangle\Big|.
\end{eqnarray}
This observable provides insight into the degree of thermalization in the system by quantifying the deviation from an equal distribution of particles between the even and odd sites. Therefore, at long times the imbalance is expected to become zero in the ergodic regime. Figure~\ref{fig:imbal} shows the evolution of the imbalance for both the regimes. When the disorder is weak, the imbalance decays, and its long time evolution saturates close to zero.
In contrast, in the strong disorder regime the imbalance remains significantly large and stays close to its initial value. This behavior clearly demonstrates the many-body localization regime where the system's initial memory is significantly retained. In this regime, the non-Hermitian ETH framework does not hold. 

This strongly supports the effectiveness of our formalism beyond the prototypical chaotic models of  quantum many-body systems that we have considered in our work, the Ginibre ensemble and SYK model. Indeed, the disorder in the HN model offers a control knob to tune between an ergodic ETH regime and a localized non-ETH regime.  
Furthermore, as discussed below, the feasibility of implementing the HN model in current experimental setups creates an opportunity for direct testing of our predictions under the presented non-Hermitian ETH framework. 

\subsubsection*{Experimental feasibility}

The time evolution of an open quantum system coupled to a large Markovian environment is governed by the Lindblad master equation~\cite{book_breuer_petruccione} 
\begin{equation}
\partial_{t}\rho = -i[H, \rho] + \sum_{m}\gamma_{m}\left[\hat{\mathcal{K}}_{m}\rho\hat{\mathcal{K}}_{m}^{\dagger}-\frac{1}{2}\{\hat{\mathcal{K}}^{\dagger}_{m}\hat{\mathcal{K}}_{m}, \rho\} \right],
\end{equation}
where $\rho$ is the density matrix of the system, $\hat{\mathcal{K}}_{m}$ are the jump operators of $m$ different channels, $\gamma_{m}$ are the jump strengths, and $[\bullet]$ ($\{\bullet\}$) denote the commutation (anti-commutation) operation. Here, the first term in the right hand side corresponds to Hermitian evolution of the system under the Hamiltonian $H$. The equation can be cast~\cite{daley_2014} as
\begin{equation}
\partial_{t}\rho = -i[H_{\rm nh}, \rho] + \sum_{m}\gamma_{m}\hat{\mathcal{K}}_{m}\rho\hat{\mathcal{K}}_{m}^{\dagger},
\end{equation}
where $H_{\rm nh} = H - \frac{i}{2}\sum_{m}\gamma_{m}\hat{\mathcal{K}}_{m}^{\dagger}\hat{\mathcal{K}}_{m}$ corresponds to a non-Hermitian Hamiltonian~\cite{zoller_1,zoller_2}. Thereby, a non-Hermitian evolution can be inferred in the limit of no-jumps. Alternatively, one can regard the evolution between consecutive jump processes as non-Hermitian when the dynamics is studied at the level of individual trajectories~\cite{molmer_1993, daley_2014}. Therefore, for suitably post-selected trajectories the system undergoes an evolution under an effective non-Hermitian Hamiltonian. As we discuss now, such a post-selection permits one to realize a Hatano--Nelson model as described above using suitably engineered dissipation in an optical lattice.  

Consider the Hermitian Hamiltonian 
\begin{equation}
   H=\sum_i^N\big[ -J\cosh{(\chi)}( \hat{c}^\dagger_i\hat{c}_{i+1}+ \hat{c}^\dagger_{i+1} \hat{c}_{i})+V \hat{n}_i \hat{n}_{i+1} + h_{i}\hat{n}_i\big]\,.
\end{equation}
This Hamiltonian describes a 1D optical lattice system loaded with ultracold atoms, which can be either spin-polarized fermions or bosons~\cite{Lewenstein_2007,Dutta_2015,Gross_2017,Tarruell_2018,Takahashi_2020}. In the case of bosonic atoms, the filling and onsite contact interaction can be tuned such that more than one occupancy per site (hardcore condition) is prohibited. The onsite interaction strength is well controlled through magnetic Feshbach resonances~\cite{chin_2010}. The nearest-neighbour interaction becomes relevant for atoms having large magnetic dipole moments~\cite{Trefzger_2011, Chomaz_2022}. In recent optical lattice experiments, the effects of the nearest-neighbour interaction have been demonstrated~\cite{Baier_2016,su_2023}. In Ref.~\cite{Baier_2016}, an optical lattice with magnetic Er atoms and small lattice spacing was considered, which allowed to tune the nearest-neighbor interaction strength from $V\approx J$ to $V\approx -2J$ by tilting the polarization axis.
The regime of $V\approx 10J$ has been achieved in a quantum gas microscopy set-up with a variable spacing optical accordion lattice and magnetic Er atoms~\cite{su_2023}. The site-dependent disorder can be engineered by manipulating the conservative optical dipole potential of the lattice either through super-imposing a speckle pattern~\cite{Billy_2008, choi_2016} or by interfering an incommensurate optical lattice potential yielding quasi-periodic disorder~\cite{Roati_2008, Schreiber_2015, Bordia_2017, Luschen_2017, Rispoli_2019, Sauerwein_2023}. The strength of the disorder potential can be tuned by controlling the intensity of the introduced light-field. In the previous references, experimental realizations of disorder strengths as large as $W\approx 10J$ and $\approx 20J$ have been reported for the quasi-periodic~\cite{Bordia_2017,Luschen_2017} and speckle disorders~\cite{choi_2016}, respectively. 

To arrive at a non-Hermitian Hamiltonian as in Eq.~\eqref{eq:ham_hnmodel}, we can consider the system to be subjected to non-local one-body loss processes governed by the jump operators
\begin{equation}
\mathcal{K}_i = \sqrt{2J\sinh{(\chi)}} (\hat{c}_i -i\hat{c}_{i+1}).
\end{equation}
Unlike on-site one-body loss process, engineering the above processes is challenging. Reference~\cite{Gong_2018} proposes its implementation through superimposing a running wave, which provides non-local Rabi coupling in a dissipative optical lattice system~\cite{Tomita_2019}. Considering experimentally feasible parameters, the strengths of the non-reciprocal hopping process can be worked out to be $J_{R} = J e^{\chi}= 1.1J$ and $J_{L}=J e^{-\chi}= 0.9J$ (see Appendix F of Ref.~\cite{Gong_2018} for further details).
Alternatively, the non-reciprocal hopping can be implemented in momentum space lattice through multi-photon Bragg transitions, flux introduction and controlled dissipation engineering~\cite{Gou_2020}.  

The considered observables, single-site occupancy and the particle imbalance, can be monitored in optical lattice systems through site resolved fluorescence spectroscopy. This is generically  implemented in quantum gas microscopy set-ups~\cite{Gross_2021}. Also, the number of atoms in odd and even sites can be read off through band mapping techniques yielding the imbalance~\cite{Schreiber_2015}. Through these techniques, many-body localization has already been investigated in modern day quantum simulators comprising cold atoms in 1D and 2D optical lattices \cite{Schreiber_2015, choi_2016, Bordia_2017, Luschen_2017, Rispoli_2019}. 

In the previous section, we presented the dynamics of the Hatano-Nelson model for parameters $V =2J$, $W = 1J$ and $20J$, and $\chi = 0.1$ yielding $J_{R}\approx 1.1J$ and $J_{L}\approx 0.9J$. These parameters are within the reach of present day experiments as outlined above. As shown in our numerical analysis presented in Figs.~\ref{fig:hatnel_numop} and \ref{fig:imbal}, the relevant signatures emerge already on timescales of the order of $Jt \sim 10$. Combined with the experimental setup and accessible probes outlined in this section, this confirms that non-Hermitian thermalization can be effectively monitored in current cold-atom experiments.


\end{document}